\documentclass[twoside,a4paper]{article}

\usepackage{a4wide}
\usepackage{anslistings}
\usepackage{algorithm,algorithmicx,algpseudocode}
\usepackage{amsmath,amssymb}
\usepackage[noblocks]{authblk}
\usepackage{booktabs}
\usepackage{fancyhdr}
\usepackage{graphicx}
\usepackage[hyperfootnotes=false,colorlinks,linkcolor=blue,urlcolor=blue,citecolor=blue]{hyperref}
\usepackage{natbib}
\usepackage{lastpage}
\usepackage{listings}

\usepackage{enumitem}
\usepackage{mathtools}
\usepackage{tikz}
\usepackage{transparent}
\usepackage{subcaption}

\title{Dune-CurvedGrid -- A Dune module for surface parametrization}
\author[1]{Simon Praetorius}
\author[2]{Florian Stenger}
\affil[1]{Technische Universit{\"a}t Dresden, Institut f{\"u}r Wissenschaftliches Rechnen,
  D-01062 Dresden, Germany (\url{simon.praetorius@tu-dresden.de}).}
\affil[2]{Technische Universit{\"a}t Dresden, Institut f{\"u}r Wissenschaftliches Rechnen,
  D-01069 Dresden, Germany (\url{florian.stenger@tu-dresden.de}).}

\newlist{types}{description}{3}
\setlist[types]{itemsep=1pt,font=\ttfamily,itemindent=0pt,leftmargin=\labelsep}

\newtheorem{prop}{Proposition}

\newcommand{\X}{\mathbf{X}}
\newcommand{\Dune}{\textsc{Dune}}
\newcommand{\Gmsh}{\textsc{Gmsh}}

\newenvironment{code++}{\begin{minipage}{\linewidth}\begin{c++}}{\end{c++}\end{minipage}}

\begin{document}
\maketitle

\begin{abstract}
  In this paper we introduce and describe an implementation of curved (surface) geometries within the \Dune\ framework for grid-based discretizations. Therefore, we employ the abstraction of geometries as element-functions bound to a grid element, and the abstraction of a grid as connectivity of elements together with a grid-function that can be localized to the elements to provide element local parametrizations of the curved surface.
\end{abstract}

\section{Introduction}
Numerical computations on curved surfaces are an important tool for studying physical phenomena in thin structures, on curved boundaries of domains, and on interfaces between two bulk regions. Those problem arise, e.g., in fluid dynamics when considering free-surface flows, in interfacial transport problems on biological membranes or fluid droplets, in mathematical geosciences, and in the physics of thin films with vanishing film thickness.
All these applications require the discretization of partial differential equations (PDEs) on the embedded surface, involving quantities like curvature, surface measures, normal vectors, and covariant derivatives that need to be available by a numerical method. An overview about some of these applications can be found in \cite{BoRe2017,NeNiPrVo2017,GL2018Analysis,JaOlRe2018,FrSc2009}.

In discretization methods based on finite elements or finite volumes one is often faced with the problem of representing the curved geometry by numerical grids and discretizing a (geometric) partial differential equation on this approximation of the actual smooth surface. One distinguishes between implicit surface representations using cut cells, level-sets or diffuse interfaces, see \cite{RaVo2006,Le2016High,OlRe2017,BuHaLaMa2018}, and explicit surface representations by triangulations and surface finite element or surface finite volume schemes. For the latter, the representation with piecewise flat elements is a widely used lowest order approximation that is easy to construct and to implement with standard software. Unfortunately, it has some drawbacks. The local flat approximation leads to vanishing curvature inside the elements and this can sometimes lead to non-converging numerical schemes, as shown in \cite{Heine2004} for the discrete mean-curvature vector and Weingarten map, in \cite{Fritz2013} for a finite-element approximation of the Ricci curvature, and in \cite{HaLaLa2019,HaPr2021Tangential} for the discretization of a surface vector Laplacian\footnote{In these three examples a stabilization technique or additional higher-order geometric knowledge allows to overcome these flat element limitations}.
The numerical error involved in a discretization of PDEs on curved surfaces depends on two properties of the discretization, the representation of the objective function and the representation of the geometry. Thus, a higher-order scheme is only possible with also a higher-order description of the surface approximation.

Better approximations of the geometry than piecewise flat are necessary. Those can be found in piecewise polynomial approximations or even exact representations of the surface if an analytical description of the surface's geometry is available. Some numerical libraries for the implementation of partial-differential equation solvers, like, e.g., \cite{dealII,Anderson2021,GeRe2009,JoReGe2014,Schoberl1997,Schoberl2014}, allow for higher-order geometry parametrization. Others do not always provide a usable interface for such non-linear geometry transformation or representation of such curved grids and often do not provide the utilities necessary to implement such simulations efficiently.

The present work tries to fill this gap for the numerical library \Dune, \cite{Dune}, a framework  for the discretization of grid-based numerical problems. This modular library is centered around the abstraction of a grid interface and the coupling to various different grid implementations for structured or unstructured grids, with or without adaptivity, sequential or parallel traversal, for volume and surface grids and additionally allowing to wrap any grid implementation to extend or modify its functionality. This concept of wrapped grids, called \emph{meta-grids} in \Dune, is the basis of our implementation. We provide a wrapper around any grid implementation in \Dune, transforming a (piecewise) flat reference grid into a curved grid using a (non-linear) geometry transformation in the local elements while preserving the grid connectivity and element numbering.

All implementations discussed in this paper can be found in individual \textsc{Dune} modules in publicly available source code repositories: \cite{DuneCurvedGrid,DuneCurvedGeometry,DuneVtk,DuneGmsh4}. The code examples and results in this manuscript are summarized and made available in the repository \cite{DuneCurvedGridExamples}.

\subsection{Initial Example}
For a motivation of the introduced functionality in \textsc{Dune-CurvedGrid}, we consider the polynomial approximation of a spherical surface by local polynomials of some given \texttt{order}.

\begin{c++}
// 1. Construct a reference grid
std::unique_ptr refGrid =
  Gmsh4Reader< FoamGrid<2,3> >::createGridFromFile("sphere.msh");

// 2. Define the geometry mapping
auto sphere = [](const auto& x) { return x / x.two_norm(); };
auto sphereGridFct = analyticDiscreteFunction(sphere, *refGrid, order);

// 3. Wrap the reference grid to build a curved grid
CurvedGrid grid{*refGrid, sphereGridFct};
\end{c++}

The curved grid is build on top of a piecewise flat reference grid provided by a \Gmsh\footnote{The \Gmsh\ format is described in \cite{GeRe2009}} file and represented by a \textsc{Dune-FoamGrid}, \cite{SaKoScFl2017}. Although this reference grid is created using the \texttt{Gmsh4Reader} which will be discussed in \autoref{sec:input_output}, the grid could also be created by any other method\footnote{For example \textsc{Dune-Grid} provides it's own \texttt{GmshReader} which supports the older file format version 2.}. The polynomial representation is given by interpolating an analytic coordinate projection, \texttt{sphere}, i.e., a simple normalization of global coordinates, locally into a polynomial space of given polynomial \texttt{order}. This is achieved by the function wrapper \texttt{AnalyticDiscreteFunction}, that provides evaluation of values and derivatives of the given function in terms of its local discrete approximation, see \autoref{sec:grid-functions}.

The library provides grid-functions that can be used to represent geometry mappings. Both, analytical and discrete representations, are implemented and thus even evolving parametrizations as solutions of PDEs or other external descriptions are possible.

Finally, the reference grid together with the coordinate mapping build the curved grid. The \texttt{CurvedGrid} fulfills the \Dune\ grid requirements and can be used instead of any other regular \Dune\ grid. It provides the geometric mappings for grid elements and for element intersections, allowing for continuous and discontinuous discretization schemes build on top of this grid.

\subsection{Structure of the Paper}
In \autoref{sec:parametric_discrete_surfaces} the mathematical foundation is laid, describing smooth surfaces and their (polynomial) approximation, the grid and its geometry transforms as well as grid-functions. This section introduces the notation and defines the objects and mappings that are implemented in the library. The subsequent \autoref{sec:geometry} introduces the interface for the geometry classes representing the local mapping of coordinates. This is followed by a section about the class interface for the \texttt{CurvedGrid} in \autoref{sec:grid} that implements the grid wrapper providing locally defined curved geometries. The description of the actual surface or its approximation requires grid-functions, introduced in \autoref{sec:grid-functions}, and projection mappings that are shown by examples in \autoref{sec:geometries}. This concludes the implementation aspects of the library. In \autoref{sec:numerical_examples} numerical validation of the implementation is given by analyzing known error bounds for geometric quantities like distance between surfaces, normal vectors, and mean curvature. This is followed by a selection of numerical examples of finite-element problems on curved surfaces.

\section{Parametric Discrete Surfaces}\label{sec:parametric_discrete_surfaces}
Let $\Gamma\subset\mathbb{R}^{m+1}$ be an oriented, connected, and smooth $m$--dimensional manifold. $\Gamma$ can be described in multiple ways, e.g., by a parametrization over a reference domain, by an implicit representation as level-set of a function, or by closest-point projection of coordinates on another manifold in a close neighborhood of $\Gamma$. All these descriptions have advantages and disadvantages and are summarized in \cite{DzEl2013}. While continuous descriptions allow to extract geometric measures and characteristics of the surface, like its metric or curvature, they are complicated to use in numerical computations. Hence, an approximation, or piecewise representation of the surface for local evaluation of quantities and data is desirable.

\subsection{Reference Geometry}
Such a representation might be given by a piecewise flat surface $\Gamma_h$, topologically equivalent to the smooth surface $\Gamma$. This reference surface is composed of finitely many regular and quasi-uniform (flat) $m$--dimensional elements of diameter $h$. The collection of these patches, typically simplices or hyper-cubes, is denoted by $\mathcal{G}_h$ and is called the grid representation of $\Gamma_h$, with
\begin{equation}
  \Gamma_h = \bigcup_{e\in\mathcal{G}_h} e
\end{equation}
where $e$ denotes an element of the grid. We assume that the patches do not overlap, i.e., for $e_1,e_2\in\mathcal{G}_h$ we have that $\operatorname{int}(e_1)\cap \operatorname{int}(e_2)=\emptyset$ and if $e_1\cap e_2=I\neq\emptyset$ and $\operatorname{dim}(I)=m-1$, it is called an intersection of $e_1$ and $e_2$ and is assumed to be a subset of an $(m-1)$--dimensional facet of $e_1$ and $e_2$, respectively.

Each element of the grid $e\in\mathcal{G}_h$ is parametrized over a reference element $\hat{e}\subset\mathbb{R}^m$ by an invertible and differentiable mapping $\mu_e: \hat{e}\to e$, called the geometry mapping of $e$. Additionally, we assume that there exists a bijective mapping $\X:\Gamma_h\to\Gamma$, such that the smooth surface can be represented by the union of (non-overlapping) mapped patches, i.e.,
\begin{equation}
  \Gamma = \bigcup_{e\in\mathcal{G}_h} \X(e)
         = \bigcup_{e\in\mathcal{G}_h} \X(\mu_e(\hat{e}))
 \eqqcolon \bigcup_{e\in\mathcal{G}_h} \X_e(\hat{e}) \,.
\end{equation}
With this property, we call $\Gamma_h$ the reference manifold or reference domain of $\Gamma$ and the family $\{\X_e\}_{e\in\mathcal{G}_h}$ its reference parametrization.

\subsection{(Higher-order) Approximations of the Manifold}
The reference manifold $\Gamma_h$ from the last section is not used directly for an approximate discretization of functions on $\Gamma$, since it does not necessarily approximate the smooth manifold well enough. It just provides a reference domain for the parametrization $\X$. For numerical computations and discretizations, we need another manifold in the proximate neighborhood of $\Gamma$. For a piecewise polynomial surface approximation, we follow the general notation of \cite{De2009}.

\begin{figure}[ht]
  \begin{center}
    \input{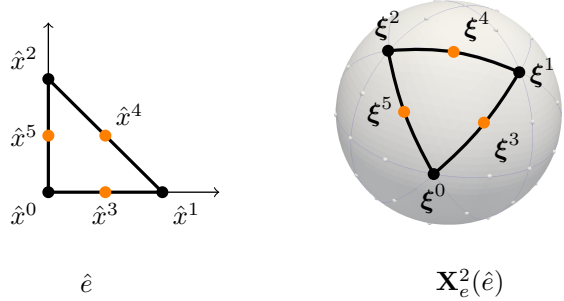}
  \end{center}
  \caption{\label{fig:parametrization}Lagrange parametrization of order $k=2$ with Lagrange nodes on
  vertices and edges.}
\end{figure}

Therefore, let $\X^k\coloneqq \mathbb{I}_h^k \X\in\mathbb{P}_k(e)$ be a $k$th-order polynomial (Lagrange) interpolation of the mapping $\X$ on the element $e$ of the reference manifold, with Lagrange nodes sitting on the smooth surface $\Gamma$. $\mathbb{P}_k(e)$ denotes the space of polynomials on $e$ of degree at most $k$ and $\mathbb{I}$ the (componentwise) Lagrange  interpolation operator. The interpolation can be expressed in terms of local basis functions in the reference element $\hat{e}$, by
\begin{equation}\label{eq:lagrange_parametrization}
  \X^k(\mu_e(\hat{x})) \coloneqq \X_e^k(\hat{x})
    = \sum_{j=1}^{n_k} \boldsymbol{\xi}^j\cdot \phi_j(\hat{x}),\quad\text{for }\hat{x}\in\hat{e}\,,
\end{equation}
with $\boldsymbol{\xi}^j\coloneqq \X_e(\hat{x}^j)\in\Gamma$ for $\{\hat{x}^j\}_{j=1\ldots n_k}$ local Lagrange nodes, the corresponding local Lagrange basis functions $\{\phi_j\}_{j=1\ldots n_k}$, and $n_k$ the number of local basis functions of $\mathbb{P}_k(\hat{e})$. The nodes and basis functions fulfill the nodal interpolation property $\phi_i(\hat{x}^j)=\delta_{ij}$. Thus, the mapped Lagrange nodes $\boldsymbol{\xi}^j$ sit on the smooth surface $\Gamma$, see \autoref{fig:parametrization} for an illustration.

Then, the $k$th-order approximation $\Gamma_h^k$ of the smooth surface $\Gamma$ can be obtained by the union of (non-overlapping) mapped elements, mapped by $\X^k$:
\begin{equation}
  \Gamma_h^k = \bigcup_{e\in\mathcal{G}_h} \X^k(e)
             = \bigcup_{e\in\mathcal{G}_h} \X^k(\mu_e(\hat{e}))
             = \bigcup_{e\in\mathcal{G}_h} \X_e^k(\hat{e}) \,.
\end{equation}

In case of $k=1$ we speak of a piecewise flat or polyhedral surface grid $\Gamma_h^1$. Since the mappings $\X_e^k$ are locally smooth, we obtain a piecewise differentiable manifold.

\subsection{The Grid, Entities, and Intersections}
The reference manifold $\Gamma_h$ is composed of elements $e$ that build the grid $\mathcal{G}_h$, or vice versa, the grid defines the manifold. The (higher-order) mapped elements and the elements mapped to the smooth manifold also form grids, namely
\begin{equation}
  \mathcal{G}_h^k \coloneqq \{\X^k(e)\}_{e\in\mathcal{G}_h}\;\text{ and }\;
  \mathcal{G}     \coloneqq \{\X(e)\}_{e\in\mathcal{G}_h}\,.
\end{equation}

Since, we assume that all manifolds $\Gamma$, $\Gamma_h$, and $\Gamma_h^k$ have the same dimension, also the grids are composed of elements $e$ of that same dimension $m$. We speak of a conforming grid if the surface is continuous and the non-empty intersection of each two elements $e_1$ and $e_2$ is an $l$--dimensional facet of both elements, called sub-entity $s\preceq e_1$ and $s\preceq e_2$. We say $s$ has co-dimension $c=m-l$. In conforming grids the sub-entities of co-dimension one are the intersections $I$ of two elements.

\begin{figure}[ht]
  \begin{center}
    \input{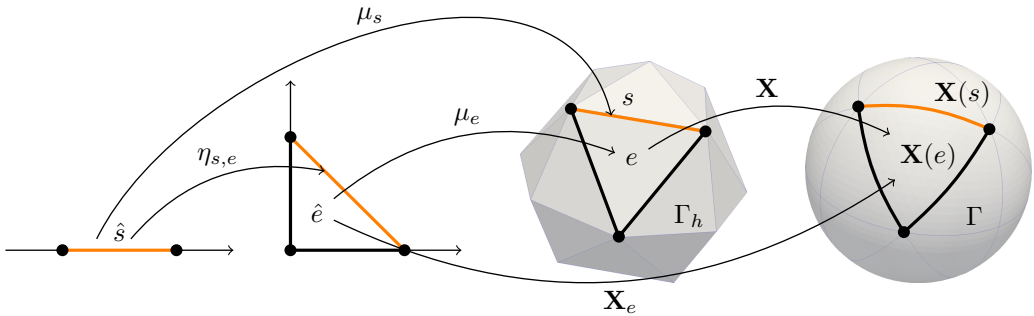}
  \end{center}
  \caption{\label{fig:mappings}Coordinate mappings $\mu$ and $\eta$ between reference element, flat element, and curved element. An additional mapping $\X_{s,e}$ can be defined by chaining of $\eta,\mu$ and $\X$, i.e. $\X_{s,e}=\X\circ\mu_e\circ\eta_{s,e}$.}
\end{figure}

Corresponding to the reference element $\hat{e}$ there is a reference element $\hat{s}\subset\mathbb{R}^l$ of the sub-entity $s$ of $e$. The relation between the geometries of $\hat{s}$ and $\hat{e}$ is given by the invertible and differentiable mapping $\eta_{s,e}:\hat{s}\to\hat{e}$, called the local-geometry mapping between sub-entity and element. With this, the parametrization of the real sub-entity $s$ is given by the chain of $\eta$ and $\mu$, i.e., $s=\mu_e(\eta_{s,e}(\hat{s}))$, see \autoref{fig:mappings} for an illustration. If this parametrization is equivalent to the direct mapping from the reference element $\hat{s}$ to $s$, i.e., if it holds $\mu_e\circ\eta_{s,e} = \mu_s$ for all sub-entities $s$ of $e$, we call the grid \emph{twist-free}, see \cite{DeNo2010}. This property is assumed to hold at least for intersections, in the following. Examples of twist-free grids are \texttt{Dune::OneDGrid}, \texttt{Dune::YaspGrid}, and \texttt{Dune::ALUGrid}.

\subsection{Grid-functions and Element-functions}
When discussing the mapping $\X$ or its polynomial variants $\X^k$, we often define it by its local representation $\X_e$ and $\X_e^k$, respectively, with $e$ the element it is defined on. Those functions, defined via their local element variant, are called \emph{grid-functions} in the  following and are directly connected to their local variant, called \emph{element-functions}. The evaluation of the grid-function in global coordinates $x\in\mathbb{R}^n$ might be an expensive operation, whereas the evaluation in the corresponding local coordinate $\hat{x}$ can be easily defined. An example is the evaluation of a discrete function by linear combination of evaluations of local basis functions inside the grid elements.

In general, we denote by $f_e:\hat{e}\to\mathcal{R}, f_e = f\circ\mu_e$ the element-function bound to the element $e$. It is associated to the grid-function $f:e\to\mathcal{R}$ with range $\mathcal{R}$. If $f$ is smooth or at least differentiable inside the element $e$, we denote by $Df:e\to L(e,\mathcal{R})$ its derivative as linear mapping, often represented as a matrix $\mathbb{R}^{\operatorname{dim}(\mathcal{R})\times n}$. The corresponding localized derivative $(Df)_e$ is then given by
\begin{equation}
  (Df)_e:\hat{e}\to L(e,\mathcal{R}),\quad (Df)_e=Df\circ\mu_e\,,
\end{equation}
with the same linear mapping in the range as the global $Df$. This notation follows \cite{EnGrMuSa2017}. If instead just a local Jacobian $D(f_e):\hat{e}\to L(\hat{e},\mathcal{R})$ is available or requested, we have the relation $D(f_e) = (Df)_e\cdot D(\mu_e)$.

Note that the geometry mapping $\X$ is a grid-function with element-function $\X_e$. Also, the parametrized geometry mappings $\X^k$ and $\X^k_e$ are grid-function and element-function, respectively. These mappings are  differentiable, by differentiating their local basis functions $\phi_j$, i.e.,
\begin{equation}\label{eq:discrete_lagrange_jacobian}
  D(\X_e^k) = \sum_j^{n_k} \boldsymbol{\xi}^j\otimes D(\phi_j)\,.
\end{equation}

In the next section we want to introduce the implementation of the element geometry mappings $\X_e$ and $\X_e^k$ and then in the subsequent section a wrapper to transform the reference grid $\mathcal{G}_h$ into $\mathcal{G}$ or $\mathcal{G}_h^k$ using the geometry mappings and its global variants $\X$ and $\X^k$ from above.

\section{\textsc{CurvedGeometry} and the \Dune\ geometry interface}\label{sec:geometry}
A \textit{Geometry} is a mapping from local coordinates in $\mathbb{R}^l$ to global coordinates in $\mathbb{R}^n$, where the local coordinates are in the coordinate system of an entity $e$ which this geometry belongs to. The entity can be an element of the grid, or a sub-entity of any co-dimension. This flexibility requires the geometry parametrization to be evaluable in different coordinate systems and also its derivatives to be available for the corresponding coordinate transformations.

Depending on how the parametrization of the geometry is given, different implementations are provided. The \texttt{ParametrizedGeometry} expects the mapping $\X_e$ and a local finite-element and constructs the local interpolation $\X_e^k$ internally, whereas the \texttt{LocalFunctionGeometry} wraps a given $\X_e$ or $\X_e^k$ directly.

\subsection{Geometry with Local Interpolation}
The first implementation, \texttt{ParametrizedGeometry}, expects only a callable function that maps entity local coordinates to global coordinates. This mapping is internally interpolated into a local finite-element space, e.g., local Lagrange functions, that allows to evaluate values and Jacobians of the parametrization from linear combinations of evaluated local basis functions and its derivatives.

\begin{c++}
// <dune/curvedgeometry/parametrizedgeometry.hh>

template <class LocalFiniteElement, int coorddim, class Traits = (...)>
class ParametrizedGeometry;
\end{c++}

The template parameters are defined by
\begin{types}
  \item[LocalFiniteElement] A class representing a local finite-element in the sense of \textsc{Dune-LocalFunctions}.
  \item[coorddim] Dimension $n\geq l$ of the global coordinates this geometry maps into.
  \item[Traits] (optional) Parameters for internal optimization of operations.
\end{types}

This geometry mapping directly corresponds to the description of the discrete geometry $\X^k$ where the input function represents the mapping $\X$ combined with the local-to-global mapping $\mu_e\circ\eta_{s,e}$. Thus, the actual input is $\X_{s,e}$. In the geometry, the local interpolation $\X_{s,e}^k$ is represented by the interpolation coefficients $\{\X_{s,e}(\hat{x}^j)\}_j$, i.e., Lagrange nodes on the surface $\Gamma$, and the set of local basis functions $\{\phi_j\}$ associated to these nodes, see \eqref{eq:lagrange_parametrization}. The Jacobian of the geometry mapping can thus be provided by evaluating the gradients of the local basis functions and its linear combination with the stored coefficients, see \eqref{eq:discrete_lagrange_jacobian}.

The \texttt{LocalFiniteElement} parameter here is the crucial input characterizing which type of local basis functions and local interpolation to use for calculating and representing the (Lagrange) nodes. \textsc{Dune-LocalFunctions} provides various implementations of local finite-elements, like Lagrange functions on all supported geometry types. The corresponding local finite-element can be obtained either by an explicit instantiation if the geometry type is known and identical for all elements, or by using a local finite-element cache. The latter provides the local finite-element of one kind for all geometry types by type-erasure or variadic visitors, see the example below.

\subsection{Geometry with Differentiable Parametrization}
The second implementation, \texttt{LocalFunctionGeometry}, expects a mapping for coordinates and additionally the Jacobian of that mapping so that the geometry Jacobian can be represented directly by the function. We expect this mapping function to be compatible with a \texttt{ElementFunction} interface, to be defined below. An element-function $f_e$ typically can be evaluated only in element local coordinates of the element $e$, denoted by $\mathbb{R}^m$, but not in codim $> 0$ entity-local coordinates, denoted by $\mathbb{R}^l$. In order to allow the geometry to be defined also for these entities, or even element intersections, this geometry implementation is parametrized additionally with a \texttt{LocalGeometry} coordinate transform $\eta$. This coordinate transform maps the entity-local coordinates to the grid-element local coordinates where the element-function can be evaluated in. Thus, the geometry mapping is a chaining $\mathbb{R}^l\to\mathbb{R}^m\to\mathbb{R}^n$ with $l \leq m \leq n$, that is, $\X = f_e\circ \eta = f\circ\mu_e\circ \eta$, where $f$ is the global grid-function associated to the element-function $f_e$ that is bound to an element $e$.

\begin{c++}
// <dune/curvedgeometry/localfunctiongeometry.hh>

template <class ElementFunction, class LocalGeometry, class Traits = (...)>
class LocalFunctionGeometry;

template <class ElementFunction, class ctype, int dim>
using ElementLocalFunctionGeometry = LocalFunctionGeometry<ElementFunction,
  DefaultLocalGeometry<ctype,dim,dim>, LocalFunctionGeometryTraits<ctype> >;
\end{c++}

This geometry is parametrized with the types \texttt{ElementFunction}, \texttt{LocalGeometry}, and \texttt{Traits} that fulfill the following requirements:
\begin{types}
  \item[LocalGeometry] represents a geometric mapping from an entity of codim \texttt{c} to the element with \texttt{codim} 0. The geometry is bound to the domain element whereas the \texttt{ElementFunction} can be bound to the range element of this geometric mapping. Thus, it is a differentiable function $\eta:\mathbb{R}^l\to\mathbb{R}^m$ with $l=m-c$ that fulfills a reduced \textsc{Dune::Geometry} concept, i.e., there is a local-to-global mapping from the coordinate system of the codim-\texttt{c} entity to the element geometry. Let $\eta$ be of type \texttt{LocalGeometry} and $\hat{x}$ of type \texttt{LocalCoordinate}, then the expression $\eta(\hat{x})\in\mathbb{R}^m$ results in a type that is the domain type of the \texttt{Localfunction}.

  The derivative of this parametrization must be accessible by evaluating the geometry method $\eta$\texttt{.jacobianTransposed($\hat{x}$)} that returns the transposed of the Jacobian of $\eta$, that is, $D^\top\eta$ with $D \eta:\mathbb{R}^l\to L(\mathbb{R}^n,\mathbb{R}^l)\cong\mathbb{R}^{n\times l}$.

  \item[ElementFunction] represents a differentiable mapping $f_e:\mathbb{R}^m\to\mathbb{R}^n$ with given derivative $(D f)_e:\mathbb{R}^m\to L(\mathbb{R}^n,\mathbb{R}^n)\cong\mathbb{R}^{n\times n}$. It is required that $f_e$ is a model of the concept \textit{Callable}, i.e., let $x$ be of type \texttt{LocalGeometry::GlobalCoordinate}, then the expression $f_e(x)$ must result in a valid type denoted by the \texttt{GlobalCoordinate} of the \texttt{LocalFunctionGeometry}. $f_e$ must be \textit{differentiable}, i.e., there must exist a function \texttt{derivative($f$)} whose return type is another model of the \textit{ElementFunction} concept. It returns the global derivative $D(f)_e$ of the grid-function $f$ associated to $f_e$.

  \item[Traits] (optional) is a class holding parameters for the implementation of the geometry, like the tolerance and iteration limit for a Newton solver implementing the global-to-local function. Additionally it allows to specify element properties that cannot be deduced from the \texttt{ElementFunction} or \texttt{LocalGeometry} directly, like the \texttt{GeometryType} of the element, if there is only one.
\end{types}

The definition of the \texttt{ElementFunction} follows the definition of localized functions in \cite{EnGrMuSa2017}. Especially, the definition of the range of the derivative of the element-function as derivative w.r.t. global coordinates is taken from there. The final \texttt{jacobianTransposed} of the geometry mapping $\X$ is thus given by the chaining $D^\top \X = D^\top \eta \cdot (D^\top \mu_e\circ \eta) \cdot (D f\circ\mu_e\circ \eta)^\top$. Note that the mappings $\X$, $\eta$, and $\mu_e$ are geometry mappings and thus provide only the transposed of their Jacobians, whereas the localized function $f_e$ provides the non-transposed Jacobian derivative. In order to provide the transposed of the final geometry Jacobian $D \X$, it needs to be a linear map that is \textit{transposible}, i.e. it can be applied in a transposed fashion to a vector, by implementing the method \texttt{mtv()}, a method representing the transposed matrix times vector multiplication operation, or it must be representable as a matrix directly.

\subsection{Examples for the Usage of Local Geometries}
The two geometry implementations can now be used directly to parametrize a surface while traversing a flat reference grid $\mathcal{G}_h$. The following examples show the wrapping of a flat geometry into a \texttt{LocalFunctionGeometry} and \texttt{ParametrizedGeometry}, respectively.

At first, we introduce a differentiable mapping $\text{\texttt{torus}}:\mathbb{R}^2\to\mathbb{R}^3$ representing a torus parametrization.

\begin{c++}
struct Torus {
  double const r1 = 2.0, r2 = 1.0;

  auto operator() (FieldVector<double,2> const& u) const {
    return FieldVector<double,3>{
      (r1 + r2*std::cos(u[0])) * std::cos(u[1]),
      (r1 + r2*std::cos(u[0])) * std::sin(u[1]),
            r2*std::sin(u[0])
    };
  }

  friend auto derivative (Torus t) {
    return [r1=t.r1,r2=t.r2](FieldVector<double,2> const& u) {
      return FieldMatrix<double,3,2>{
        {-r2*std::sin(u[0])*std::cos(u[1]),-(r1 + r2*std::cos(u[0]))*std::sin(u[1])},
        {-r2*std::sin(u[0])*std::sin(u[1]), (r1 + r2*std::cos(u[0]))*std::cos(u[1])},
        { r2*std::cos(u[0]),                0.0}
      };
    };
  }
};
\end{c++}

Second, we have to define a reference grid that provides the actual elements $e$, its topological connectivity, and the mapping $\X$ for the curved geometry. The reference grid is a simple structured grid \texttt{YaspGrid}.

\begin{c++}
// Construct a reference grid as a (periodic) 2d rectangular domain
auto refGrid = YaspGrid<2>{{2*M_PI,2*M_PI}, {8,16}, std::bitset<2>("11")};

// Define the geometry mapping
auto torus = Torus{};
auto torusGridFct = analyticGridFunction<YaspGrid<2>>(torus);
\end{c++}

The mapping \texttt{torus} is transformed into a differentiable grid-function \texttt{torusGridFct} using the wrapper \texttt{AnalyticGridFunction}, see \autoref{sec:grid-functions}, that is provided by the library.

For the construction of a \texttt{LocalFunctionGeometry}, we have to provide an element-function of that grid-function and a \texttt{LocalGeometry}. In case of wrapping the grid element geometry, i.e., \texttt{codim} is zero, this local-geometry mapping is the identity. An efficient implementation is given by the class \texttt{DefaultLocalGeometry}.

\begin{c++}
// Define an element-function from the grid-function
auto torusElemFct = localFunction(torusGridFct);
auto localGeometry = DefaultLocalGeometry<double,2,2>{};

// traverse the reference grid
for (const auto& e : elements(refGrid.leafGridView()))
{
  // bind the element-function to the grid element
  torusElemFct.bind(e);

  // construct the LocalFunctionGeometry from ElementFunction and LocalGeometry
  auto localFctGeometry
    = LocalFunctionGeometry{e.type(), torusElemFct, localGeometry};

  // (optionally) unbind from the element, i.e., free memory and unset variables
  torusElemFct.unbind();
}
\end{c++}

An element-function must be bound to an element (and optionally unbound at the end of usage). The type supports class-template argument deduction and if the geometry is constructed on the grid element, the \texttt{LocalGeometry} argument can even be omitted, defaulting to \texttt{DefaultLocalGeometry} in this case:
\begin{c++}
auto localFctGeometry = LocalFunctionGeometry{referenceElement(e), torusElemFct};
\end{c++}
Note, it is necessary to pass the element type as \texttt{Dune::ReferenceElement}, in order to allow the deduction of the element dimension.

Similarly, we can construct a \texttt{ParametrizedGeometry} (parametrized by Lagrange local finite-elements) by using the \texttt{torus} function from above. Therefore, we have to either use the element-function wrapper or have to construct the local-to-global mapping from reference element coordinates to element coordinates directly.

A local finite-element can be provided by using a local finite-element cache, or by explicit instantiation. Both variants are shown in the example below.

\begin{c++}
// <dune/localfunctions/lagrange/lagrangelfecache.hh>
// <dune/localfunctions/lagrange/lagrangecube.hh>
...
LagrangeLocalFiniteElementCache<double, double, 2, order> lfeCache;

// traverse the reference grid
for (const auto& e : elements(refGrid.leafGridView()))
{
  // projection from local coordinates
  auto X_e = [&torus,geo=e.geometry()](const auto& local) {
    return torus(geo.global(local));
  };

  // construct the ParametrizedGeometry from lfe cache
  auto curvedGeometry = ParametrizedGeometry{e.type(), lfeCache.get(e.type()), X_e};

  // construct the ParametrizedGeometry from local finite-element
  auto lfe = LagrangeCubeLocalFiniteElement<double, double, 2, order>{};
  auto curvedGeometry2 = ParametrizedGeometry{e.type(), lfe, X_e};
}
\end{c++}

\section{\textsc{CurvedGrid} and the \Dune\ grid interface}\label{sec:grid}
Instead of wrapping the geometries manually while traversing the flat reference grid, the whole grid can be wrapped. This allows to return the wrapped geometry directly in a call to \texttt{e.geometry()} instead of the flat element geometry while preserving the grid topology and element connectivity given by the wrapped reference grid. The library provides such a grid wrapper with the class \texttt{CurvedGrid} which is an implementation of $\mathcal{G}_h^k$ or $\mathcal{G}$, depending on the element parametrization provided.

The class signature is given by

\begin{c++}
// <dune/curvedgrid/grid.hh>

template <class RefGrid, class GridFunction, bool useInterpolation = false>
class CurvedGrid;

// constructor using Lagrange geometry interpolation
template <class RefGrid, class GridFunction>
CurvedGrid<RefGrid,GridFunction,true>
  ::CurvedGrid(const RefGrid&, const GridFunction&, int order);

// constructor using LocalFunctionGeometry
template <class RefGrid, class GridFunction>
CurvedGrid<RefGrid,GridFunction,false>
  ::CurvedGrid(const RefGrid&, const GridFunction&);
\end{c++}

with template parameters
\begin{types}
  \item[RefGrid] The reference grid the curved grid is based on
  \item[GridFunction] the type of a grid-function associated with a reference grid
  \item[useInterpolation] (optional) if true, uses Lagrange \texttt{ParametrizedGeometry}, otherwise construct a geometry of type \texttt{LocalFunctionGeometry}
\end{types}

This class allows to locally construct both the \texttt{LocalFunctionGeometry} and the \texttt{ParametrizedGeometry}, depending on the properties of the grid-function passed to the grid wrapper and the \texttt{useInterpolation} parameter given to the grid. If the latter is \texttt{true}, it is assumed that a local interpolation should be constructed of the passed grid-function and thus a \texttt{ParametrizedGeometry} with Lagrange local finite-element is used as local geometry parametrization. Otherwise, if \texttt{useInterpolation} is \texttt{false} and the grid-function is locally differentiable, a \texttt{LocalFunctionGeometry} is used.

\subsection{Examples for the Usage of the Grid Wrapper}
In the following examples we construct both a wrapper using the \texttt{ParametrizedGeometry} and the \texttt{LocalFunctionGeometry}.

At first, we construct the grid by wrapping a callable representing the geometry mapping $\X$.

\begin{c++}
// Construct a reference grid
auto refGrid = Gmsh4Reader< AlbertaGrid<2,3> >::createGridFromFile("sphere.msh");

// Define the geometry mapping
auto sphere = [](const auto& x) { return x / x.two_norm(); };

// Wrap the reference grid to build a curved grid
CurvedGrid grid{*refGrid, sphere, order};
\end{c++}

In this example, a grid-function is automatically constructed from the callable \texttt{sphere} as an instance of \texttt{AnalyticGridFunction}. If class template-argument deduction from \texttt{C++17} cannot be used, a generator function \texttt{curvedSurfaceGrid(*refGrid, sphere)} is provided.

In the second example we construct a grid-function first which either uses a local interpolation internally, or is given as a differentiable function as above.

\begin{c++}
// Define a discrete grid-function on the reference grid
auto gridFct = discreteGridViewFunction<3>(refGrid->leafGridView(), order);

// Interpolate the parametrization into the grid-function
Functions::interpolate(gridFct.basis(), gridFct.coefficients(), sphere);

// Wrap the reference grid to build a curved grid
CurvedGrid grid{*refGrid, gridFct};
\end{c++}

Here, in the example we use a generic discrete function with range type \texttt{FieldVector<double,3>}, that is represented by a \textsc{Dune-Functions} global basis.
\begin{c++}
power<3>(lagrange(order), blockedInterleaved())
\end{c++}
This represents a product basis (factory) composed of three times the same Lagrange basis with an ordering of the global indices in groups of the components.
The basis is stored together with a coefficient vector inside the \texttt{DiscreteGridViewFunction}, see \autoref{sec:grid-functions}. Note that this grid-function is associated to a \texttt{GridView} and not the whole grid, since the global basis is bound to a \texttt{GridView}. Each time the (reference) grid changes, e.g., by local refinement or parallel load balancing, the grid-function must be updated as well, using
\begin{c++}
gridFct.update(refGrid->leafGridView());
\end{c++}

\section{Grid-functions and Parametrizations}\label{sec:grid-functions}
In order to construct the \texttt{ParametrizedGeometry} or \texttt{LocalFunctionGeometry} and thus the curved grid, parametrizations in form of mappings $\X_e$ or element-functions must be provided. Various grid-functions are implemented in \textsc{Dune-CurvedGrid} to simplify the construction and to act as reference implementations:
\begin{types}
  \item[AnalyticGridFunction] Implementation of a grid-function that can be bound to any entity in the grid given by a \emph{Callable}, mapping global coordinates of $\Gamma_h$ to a range type. This range type defines the global coordinates in the curved geometry. If the callable is \emph{differentiable} so is the grid-function. It can thus be used in the \texttt{LocalFunctionGeometry}.

  The \texttt{AnalyticGridFunction} can be constructed by

\begin{c++}
// <dune/curvedgrid/gridfunctions/analyticgridfunction.hh>

template <class Grid, class Functor>
class AnalyticGridFunction;

template <class Grid, class Functor>
auto analyticGridFunction (Functor&& functor)
  -> AnalyticGridFunction<Grid, std::decay_t<Functor>>;
\end{c++}

  \item[AnalyticDiscreteFunction] Similarly to \texttt{AnalyticGridFunction} this grid-function is constructed from a \emph{Callable}, mapping global coordinates to global coordinates, but the mapping is locally interpolated by means of a Lagrange basis. Thus, this grid-function does not represent an exact geometry but an approximation. Moreover, it provides derivatives by differentiating the local basis functions, see \eqref{eq:discrete_lagrange_jacobian}. It can be used to parametrize \texttt{LocalFunctionGeometry}.

  The \texttt{AnalyticDiscreteFunction} can be constructed by

\begin{c++}
// <dune/curvedgrid/gridfunctions/analyticdiscretefunction.hh>

template <class Grid, class Functor, int order = -1>
class AnalyticDiscreteFunction;

template <class Functor, class Grid>
auto analyticDiscreteFunction (Functor&& functor, const Grid&, int order)
  -> AnalyticDiscreteFunction<Grid, std::decay_t<Functor>>;

template <int order, class Functor, class Grid>
auto analyticDiscreteFunction (Functor&& functor, const Grid&)
  -> AnalyticDiscreteFunction<Grid, std::decay_t<Functor>, order>;
\end{c++}

  Note that this grid-function requires \textsc{Dune-Functions} as module dependency.

  \item[DiscreteGridViewFunction] This grid-function is restricted to a specific \texttt{GridView} of the grid and is build by a set of global basis functions and a coefficient vector, both stored inside this grid-function. It can be used as \texttt{LocalFunctionGeometry}, since the basis functions provide a derivative and thus the grid-function is differentiable. Additionally, the coefficient vector, i.e., the vector of Lagrange nodes on the smooth surface $\Gamma$, can be modified and thus evolving grids can be parametrized easily.

  The \texttt{DiscreteGridViewFunction} can be constructed by

\begin{c++}
// <dune/curvedgrid/gridfunctions/discretegridviewfunction.hh>

template <class GridView,
          int components = GridView::dimensionworld,
          int ORDER = -1,
          class T = double>
class DiscreteGridViewFunction;

template <int components, int ORDER = -1, class T = double, class GridView>
auto discreteGridViewFunction (const GridView& gridView, int order = ORDER)
  -> DiscreteGridViewFunction<GridView, components, ORDER, T>;
\end{c++}
Note, the template parameter \texttt{ORDER == -1} means, the polynomial order must be given as constructor argument or as final function argument in \texttt{discreteGridViewFunction}. Otherwise, if \texttt{ORDER >= 0}, a static polynomial order is
considered. In case the template parameter and the function argument are both negative, an error is
thrown.

This grid-function is not as general as the \texttt{DiscreteGlobalBasisFunction} of \textsc{Dune-Functions}, i.e., the range type is fixed to \texttt{FieldVector<T,components>} and the global basis is explicitly defined as \texttt{power<components>(lagrange(order), blockedInterleaved())}, see \cite{EnGrMuSa2018}. But it defines the necessary derivative and is implemented as a grid-function that includes the coefficient vector. Note that this grid-function requires \textsc{Dune-Functions} as module dependency.
\end{types}

\section{Geometries}\label{sec:geometries}
In order to test PDE discretizations or in benchmarks, geometry parametrizations for simple shapes must be available. A common example is the sphere parametrization used in all the examples above.  But, additionally, shapes with less symmetry might be of interest for benchmarks and numerical validation. We have implemented the geometries of the sphere, ellipsoid, and torus as simple parametrizable shapes. Those three geometries can be obtained by

\begin{c++}
// <dune/curvedgrid/geometries/sphere.hh>

template <int dim, class ctype = double>
class SphereProjection;

// sphere radius r
template <class Grid, class T>
auto sphereGridFunction (T r)
{
  auto sphere = SphereProjection<Grid::dimensionworld,T>{r};
  return analyticGridFunction<Grid>(sphere);
}
\end{c++}

for the sphere parametrization.

\begin{c++}
// <dune/curvedgrid/geometries/ellipsoid.hh>

template <class ctype = double>
class EllipsoidProjection;

// major axis a, b, and c
template <class Grid, class T>
auto ellipsoidGridFunction (T a, T b, T c)
{
  auto ellipsoid = EllipsoidProjection<T>{a,b,c};
  return analyticGridFunction<Grid>(ellipsoid);
}
\end{c++}

for the ellipsoid parametrization.

\begin{c++}
// <dune/curvedgrid/geometries/torus.hh>

template <class ctype = double>
class TorusProjection;

// Outer radius R and inner radius r
template <class Grid, class T>
auto torusGridFunction (T R, T r)
{
  auto torus = TorusProjection<T>{R,r};
  return analyticGridFunction<Grid>(torus);
}
\end{c++}

for the torus parametrization.

For all three a projection and a corresponding grid-function that just wraps the callable into an \texttt{AnalyticGridFunction} is provided.

Additionally, two geometry parametrizations are implemented for an explicit or implicit surface representation. That is, a representation as high-resolution (piecewise) flat grid, or as zero-level set of an implicit function. For both representations the corresponding coordinate projection is implemented which is required to construct the grid-function for the curved geometries.

\subsection{Projection to High-Resolution Surface Grid}
The explicit surface representation is based on a flat grid that approximates the smooth surface with higher-resolution than the target grid we want to construct from a reference grid. This would allow to run simulations on a coarse grid, while the surface is only given by a very fine grid. Additionally, it allows to construct higher-order geometries for a surface that is given only with piecewise flat geometries.

\begin{c++}
// <dune/curvedgrid/geometries/explicitsurface.hh>

template <class ctype = double>
class ExplicitSurfaceProjection;

// Constructor with grid and an option to activate caching
template <class ctype>
  template <class Grid>
ExplicitSurfaceProjection<ctype>
  ::ExplicitSurfaceProjection (const Grid& grid, bool cached = true);
\end{c++}

The input to construct the \texttt{ExplicitSurfaceProjection} is a (surface) grid representing the high-resolution (piecewise) flat surface. Internally, the vertices of the grid are stored in a fast search tree, a KDTree implementation based on \cite{Nanoflann2014} which supports nearest neighbor search. Each time the projection is evaluated for a global coordinate $x$ the closest vertex in the high-resolution surface grid is searched. Afterwards the adjacent elements are considered. For each of them the closest point to $x$ is determined by orthogonal projection. The closest found point is then returned.

This approach only works well if the high-resolution surface grid is of sufficient quality, i.e., no overly acute-angled elements occur. Otherwise the adjacent elements of the closest vertex don't need to contain the actual closest point to $x$. When choosing a method for creating the grid a Delaunay-triangulation, for instance, is a good candidate. Grids not fulfilling the element-quality condition can typically be adapted without loosing their important features by using the meshing tool \texttt{meshconv}, \cite{Meshconv2020}.

The decision to only consider the elements adjacent to the unique closest vertex is a compromise made for performance-reasons. By extending the nearest neighbor search in the KDTree to several closest vertices and considering adjacent elements to all of them one could allow for lower quality surface grids at the cost of sacrificing performance.

\begin{figure}[ht]
  \begin{center}
    \input{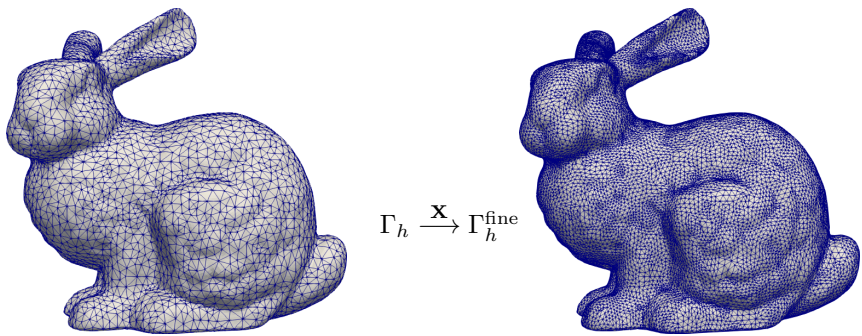}
  \end{center}
  \caption{\label{fig:bunnies}Coarse grid (left) used as reference domain $\Gamma_h$ for parametrization
  with closest-point projection to fine grid $\Gamma_h^\text{fine}$ (right) of Stanford-bunny geometry. The Coarse grid is
  obtained by feature-preserving coarsening of the fine-grid, see \cite{Meshconv2020,VaCh2004}.}
\end{figure}

\subsection{Projection to Zero-Level set}
If the surface is given implicitly as the zero-level set of a higher-order function, the closest-point projection must be calculated iteratively using either a Newton-method or a fixed-point iteration. The essential property that is used in these algorithms, is that the normal vector of the surface in a proximate neighborhood is given by the normalized gradient of the implicit function.

Thus, the input to the \texttt{ImplicitSurfaceProjection} is a differentiable function $\psi$ providing the surface as $\Gamma=\{x\in\mathbb{R}^n\,:\,\psi(x)=0\}$ with $n_\Gamma(x) = \nabla\psi(x) / \|\nabla\psi(x)\|$ at $x\in\Gamma$.

Given an initial guess for the projected point $x_0$, the authors of \cite{Persson2004,Nitschke2014} describe a scheme to iteratively compute better guesses of the projection of $x_0$ to $\Gamma$ by approximating the closest-point property $\X(x) = x - d(x)n_\Gamma(\X(x))$ with a representation of the approximate distance $d(x)\approx\psi(x)/\|\nabla\psi(x)\|$ and the normal vector representation from above:
\begin{equation}\label{eq:simple_implicit_projection}
  x_{i+1} = x_i - \nabla\psi(x_i) \frac{\psi(x_i)}{\|\nabla\psi(x_i)\|^2},\quad
  \widehat{\text{err}}_i = \frac{|\psi(x_i)|}{\|\nabla\psi(x_i)\|}\,.
\end{equation}
This scheme applies this relation iteratively, eventually converging to a point on $\Gamma$ near the closest-point $\X(x)$. It is implemented in the class

\begin{c++}
// <dune/curvedgrid/geometries/implicitsurface.hh>

template <class Functor>
class SimpleImplicitSurfaceProjection;

template <class Functor>
SimpleImplicitSurfaceProjection<Functor>
  ::SimpleImplicitSurfaceProjection (const Functor& psi, int maxIter = 10);
\end{c++}

where the maximal number of iterations in the iterative scheme is given by \texttt{maxIter}.

An improved version of that scheme that takes the iterate from the simple scheme as initial approximation of the closest-point projection and uses this point to get a better estimate for the actual distance of $x$ to $\Gamma$ is proposed in \cite{DeDz2007}:
\begin{align}
  \tilde{x}_{i+1} &= x_i - \nabla\psi(x_i) \frac{\psi(x_i)}{\|\nabla\psi(x_i)\|^2},\quad
  \text{dist} = \operatorname{sign}(\psi(x_0))\|\tilde{x}_{i+1}-x_0\| \notag \\
  x_{i+1} &= x_0 - \nabla\psi(\tilde{x}_{i+1}) \frac{\text{dist}}{\|\nabla\psi(\tilde{x}_{i+1})\|},\quad
  \text{err}_i = \widehat{\text{err}}_i + \left\|\frac{\nabla\psi(x_i)}{\|\nabla\psi(x_i)\|}\pm \frac{(x_i-x_0)}{\|x_i-x_0\|}\right\|
  \label{eq:implicit_projection}
\end{align}

The computational demand is higher than in the simple scheme, but it converges to the actual closest point on $\Gamma$. This scheme is implemented in the class

\begin{c++}
// <dune/curvedgrid/geometries/implicitsurface.hh>

template <class Functor>
class ImplicitSurfaceProjection;

template <class Functor>
ImplicitSurfaceProjection<Functor>
  ::ImplicitSurfaceProjection (const Functor& psi, int maxIter = 10);
\end{c++}

\subsubsection{Example of an application of the iterative scheme}
We consider a surface with genus two, given by the function
\[
  \psi(x,y,z) = 2y(y^{2}-3x^{2})(1-z^{2})+(x^{2}+y^{2})^{2}-(9z^{2}-1)(1-z^{2})\,.
\]

A reference surface can be obtained by extracting the zero-level set contour, e.g., by using the tool ParaView, see \cite{ParaView2005}, by a surface Delaunay triangulation combined with a surface projection, see \cite{PeSt2004,Persson2004}, or by reconstructing the implicitly defined surface using some fast marching algorithm, see \cite{EnNu2017}. We followed the first approach, combined with a coarsening of the obtained surface grid using \cite{Meshconv2020}, see also \cite{VaCh2004} for a similar approach.

Applying the implicit projection methods from above to such a coarse reference grid results in very fast convergence of both schemes to a machine epsilon tolerance. If no closest-point property is explicitly required, the simple iterative scheme \eqref{eq:simple_implicit_projection} performs faster than the full closest-point iterative projection scheme \eqref{eq:implicit_projection}.

\begin{figure}[ht]
  \centering
  \begin{subfigure}[b]{0.3\textwidth}
      \centering
      \begin{minipage}{.8\textwidth}\centering
        \begin{tabular}{l|l}
          \toprule
          scheme & iter. \\
          \hline
          \eqref{eq:simple_implicit_projection} & $4$ \\
          \hline
          \eqref{eq:implicit_projection} & $42$ \\
          \bottomrule
        \end{tabular}
      \end{minipage}
      \caption{Iteration counts}
      \label{fig:genus2:iterations}
  \end{subfigure}
  \hfill
  \begin{subfigure}[b]{0.3\textwidth}
      \centering
      \includegraphics[width=.9\textwidth]{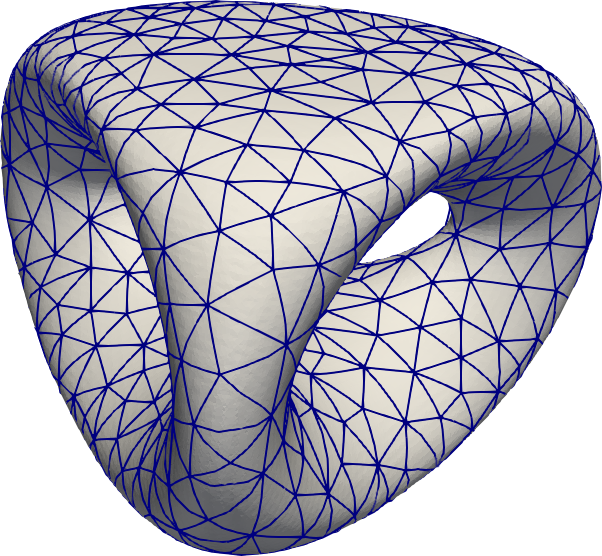}
      \caption{Grid view}
      \label{fig:genus2:side-view}
  \end{subfigure}
  \hfill
  \begin{subfigure}[b]{0.3\textwidth}
      \centering
      \includegraphics[width=\textwidth]{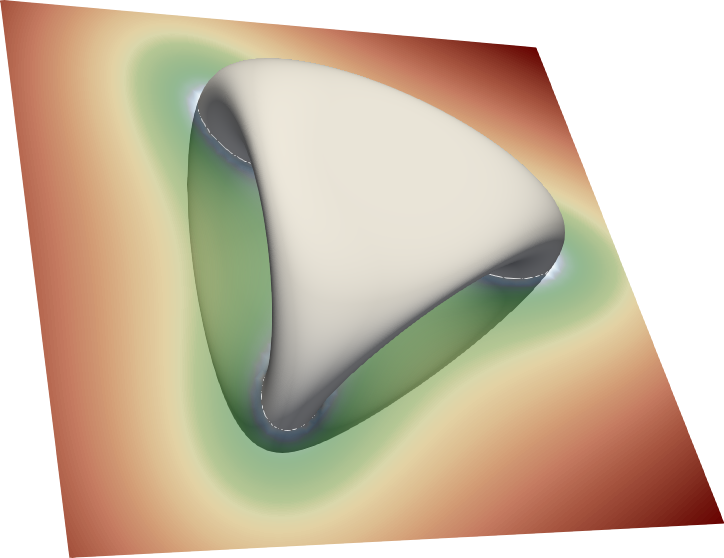}
      \caption{Levelset view}
      \label{fig:genus2:surface}
  \end{subfigure}
  \caption{\label{fig:genus2}Surface extracted from the implicit description as zero-level set of $\psi$, using an implicit projection method for the higher-order surface approximation. The colored plane illustrates a cut through the function $\psi$. The table in (a) shows the maximal number of iterations $i$ necessary to reach a pointwise tolerance $e_i < \sqrt{\epsilon}\approx 1.5\cdot 10^{-8}$ with $\epsilon$ the machine epsilon of \texttt{double} floating-point numbers of either the error $e_i=\widehat{\text{err}}_i$ that approximates the distance of $x_i$ to the surface for all projected points $x$, or the error $e_i=\text{err}_i$ in the closest-point property of the iterates.}
\end{figure}

\section{Numerical examples}\label{sec:numerical_examples}
In order to verify the implementation and to test different geometric representations, we have first analyzed the difference between the discrete surface and the continuous surface. This numerical verification considers the difference between the smooth surface quantities, the closest-point projection $\X$, the surface normal $n$, and the mean curvature $H$, in the $L^\infty(\Gamma_h)$ norm. In \cite{De2009,HaLaLa2019}, upon many others, the following estimates are shown:
\begin{prop}
  For $h$ small enough, we have the estimates 
  \begin{align}\label{eq:geometry_errors}
    \|\X - \X^k\|_{L^\infty(\Gamma_h)} \leq C h^{k+1}, \quad
    \|n\circ \X - n_h^k\|_{L^\infty(\Gamma_h^k)} \leq C h^k, \quad
    \|H\circ \X - H_h^k\|_{L^\infty(e)} \leq C h^{k-1}
  \end{align}
  for $e\in\mathcal{G}_h^k$, with $C$ a generic constant independent of the mesh parameter $h$.
\end{prop}

\begin{figure}[ht]
  \begin{center}
  \input{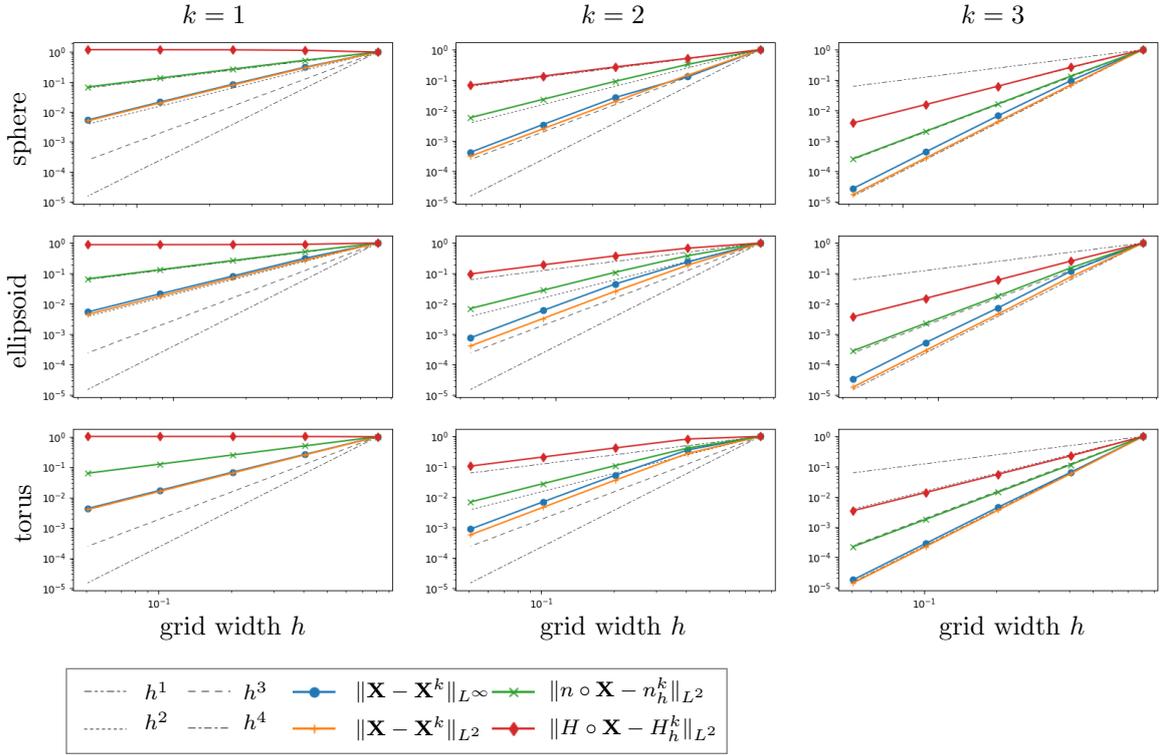}
  \end{center}
  \caption{\label{fig:convergence}Geometric error norms, normalized by error of largest grid-size, evaluated for three different geometries, the unit sphere, an ellipsoid, and a torus, for $k=1, 2$, and $3$. In various dashed lines, the ideal convergence lines $h^p$ are shown.}
\end{figure}

We show for three smooth geometries, the unit sphere, an ellipsoid with major axis $(1, 1.25, 0.75)$, and torus with the two radii $(2, 1)$, the convergence in the $L^2$-norm on the discrete surface that follows from the assertion in the $L^\infty$-norm. Therefore, we have first created a reference grid, then interpolated the surface parametrization into the element geometries with order $k$ and finally, evaluated the three quantities by iterating over the reference surface. The error norms are shown in \autoref{fig:convergence}. Note, the $L^\infty$-norm is only approximated by computing the maximum in all quadrature points of the elements that are also used for computing th $L^2$-norms.

\subsection{Surface Vector Helmholtz equation}\label{sec:surface-vector-helmholtz-eq}
We consider the vector Helmholtz equation for a test of the surface parametrization. The corresponding intrinsic surface formulation reads: Find the tangential-vector field
$u\in H^1_\text{tan}(\Gamma, T\Gamma)$ such that
\begin{equation}\label{eq:vector_helmholtz}
  \big(\nabla_\Gamma u,\,\nabla_\Gamma v\big)_\Gamma + \big(u,\,v\big)_\Gamma = \big(f,\,v\big)_\Gamma\quad\forall v\in H^1_\text{tan}(\Gamma, T\Gamma)\,,
\end{equation}
with $\nabla_\Gamma$ the covariant derivative of the vector fields, $(\cdot,\cdot)_\Gamma$ the generic $L^2$-inner product on $\Gamma$, and $f$ a given tangential vector field.

For the discretization of this equation we follow the ideas of \cite{NeNiPrVo2017,NeNiVo2019,HaLaLa2019,GrJaOlRe2018,JaRe2019} and represent the vector field in an embedding space -- in this case the Euclidean space $T\Gamma\cong\mathbb{R}^3$ -- and transform the corresponding covariant derivatives into the embedding space. By allowing the vector field to also have non-tangential components, the equation can be written as a coupled system of scalar-valued equations with a penalization term to enforce tangentiality: Find the vector field $u\in [H^1(\Gamma,\mathbb{R})]^3$ such that
\begin{equation}\label{eq:embedded_vector_helmholtz}
  \big(\nabla_\Gamma P u,\,\nabla_\Gamma P v\big)_\Gamma + \big(P u,\,P v\big)_\Gamma
    + \omega\big(n\cdot u,\,n\cdot v\big)_\Gamma = \big(f,\,P v\big)_\Gamma\quad\forall v\in [H^1_\text{tan}(\Gamma, \mathbb{R})]^3\,,
\end{equation}
with $\omega \gg 0$ a penalization parameter, $P=\text{Id}-n\otimes n$ the tangential projection operator w.r.t. the surface normal vector $n$. For extended vector fields $u$, the surface covariant derivative can be expressed in terms of the Euclidean derivative $\nabla$ in the ambient space, by $\nabla_\Gamma P u = P\nabla (P u) P = P(u\otimes \nabla_\Gamma) - (n\otimes \nabla_\Gamma)(n\cdot u)$. The expression $u\otimes\nabla_\Gamma$ means the componentwise surface gradient.

In order to discretize this equation, we approximate $\Gamma$ by $\Gamma_h^k$ and $H^1(\Gamma,\mathbb{R})$ by $V_{h,k}^r$, the Lagrange finite-element space of order $r$, given by
\[
  V_{h,k}^r\coloneqq \big\{v\in C^0(\Gamma_h^k)\,:\,v\circ\X_e^k\in\mathbb{P}_r\,\forall e\in\mathcal{G}_h\big\}
\]
Following the analysis of \cite{HaLaLa2019} and \cite{JaRe2019,GrJaOlRe2018} the normal vector involved in the covariant derivative and mass-matrix term can have the same approximation order as the geometry, but the normals involved in the penalty term should be at least one order better. We denote this ``better'' normal by $\tilde{n}$. Additionally, the scaling for the penalization factor should be of order $h^{-2}$, thus we take $\omega = \beta h^{-2}$ with $\beta = 10$ in the numerical experiments below. Note, in \cite{HaPr2021Tangential} it is shown that when interested in the tangential part of the solution only, a better choice for the approximation is $\tilde{n}\equiv n_h$ the normal vector of the discrete surface with $\omega\simeq h^{-1}$.

The resulting discrete variational formulation reads: Find the vector field $u_h\in [V_{h,k}^r]^3$ such that
\begin{equation}\label{eq:discretized_vector_helmholtz}
  \big(\nabla_{\Gamma_h^k} P_h u_h,\,\nabla_{\Gamma_h^k} P_h v_h\big)_{\Gamma_h^k} + \big(P_h u_h,\,P_h v_h\big)_{\Gamma_h^k}
    + \omega\big(\tilde{n}_h\cdot u_h,\,\tilde{n}_h\cdot v_h\big)_{\Gamma_h^k} = \big(f\circ\X^k,\,P_h v_h\big)_{\Gamma_h^k}\quad\forall v\in [V_{h,k}^r]^3\,.
\end{equation}
where the inner product and derivatives have to be understood elementwise and locally, i.e., $(A,B)_{\Gamma_h^k} = \sum_{e\in\mathcal{G}_h^k}\int_e A : B\,\text{d}\Gamma$. The challenge hereby is to integrate over the parametrized geometries and to provide normal vectors of differing approximation order. Since we have \eqref{eq:geometry_errors}, the higher-order normals can be obtained by constructing a local geometry of parametrization order $k+1$.

\begin{c++}
template <int order>
using LFE_t = LagrangeSimplexLocalFiniteElement<double,double,2,order>;

// traverse the reference grid
for (const auto& e : elements(refGrid->leafGridView()))
{
  // projection from local coordinates
  auto X_e = [&sphere,geo=e.geometry()](const auto& local) {
    return sphere(geo.global(local));
  };

  // construct the CurvedGeometries from local parametrization
  ParametrizedGeometry geometry(e.type(), LFE_t<k>{}, X_e);
  ParametrizedGeometry higherOrderGeometry(e.type(), LFE_t<k+1>{}, X_e);

  const auto& quadRule = QuadratureRules<double,2>::rule(e.type(), quad_order);
  for (const auto& qp : quadRule) {
    // integration element
    double dS = geometry.integrationElement(qp.position()) * qp.weight();
    // surface normal n_h
    auto nh = geometry.normal(qp.position());
    // higher-order surface normal n~_h
    auto nh2 = higherOrderGeometry.normal(qp.position());

    // ...
  }
}
\end{c++}

Thus, instead of traversing the \texttt{CurvedGrid}, one could iterate over the reference grid $\mathcal{G}_h$ instead and locally construct the curved geometries of order $k$ and $k+1$ to obtain the surface element and the surface normal vectors.

\subsubsection{Vector Fields on Spherical Geometry}
Following the example in \cite{NeNiVo2019} we define an exact solution $u^\ast \coloneqq \operatorname{rot}_n(x y z)$ of \eqref{eq:vector_helmholtz} and construct\footnote{The corresponding symbolic computations are done using \texttt{sympy}.} the corresponding load-vector function $f \coloneqq -\operatorname{div}_\Gamma\nabla_\Gamma u^\ast + u^\ast$. A coarse grid of the sphere is explicitly provided using a \Gmsh\ mesh with nearly equal sized elements.

In the following numerical test the geometry parametrization and function parametrization take the same polynomial order, i.e., $r=k$.

\begin{table}[ht]
  \begin{center}
  \begin{tabular}{l|l|l|l||l|l||l|l}
    \toprule
    level	& grid width $h$      & error ($k=1$)         & eoc     & error ($k=2$)         & eoc     & error ($k=3$)         & eoc \\
    \hline
    $0$ & $7.8462\cdot 10^{-1}$ & $3.0270\cdot 10^{-1}$ & ---     & $2.3687\cdot 10^{-2}$ & ---     & $1.7959\cdot 10^{-2}$ & ---     \\
    $1$ & $3.9937\cdot 10^{-1}$ & $9.2247\cdot 10^{-2}$ & $1.809$ & $3.7851\cdot 10^{-3}$ & $2.713$ & $1.1312\cdot 10^{-3}$ & $4.094$ \\
    $2$ & $2.0496\cdot 10^{-1}$ & $2.4538\cdot 10^{-2}$ & $2.000$ & $4.7189\cdot 10^{-4}$ & $3.121$ & $7.3444\cdot 10^{-5}$ & $4.099$ \\
    $3$ & $1.0322\cdot 10^{-1}$ & $6.2489\cdot 10^{-3}$ & $1.998$ & $5.6328\cdot 10^{-5}$ & $3.099$ & $4.6436\cdot 10^{-6}$ & $4.025$ \\
    $4$ & $5.1721\cdot 10^{-2}$ & $1.5703\cdot 10^{-3}$ & $2.000$ & $6.7794\cdot 10^{-6}$ & $3.064$ & $2.9136\cdot 10^{-7}$ & $4.007$ \\
    $5$ & $2.5878\cdot 10^{-2}$ & $3.9311\cdot 10^{-4}$ & $2.000$ & $8.2829\cdot 10^{-7}$ & $3.036$ & $1.8234\cdot 10^{-8}$ & $4.002$ \\
    \bottomrule
  \end{tabular}
  \end{center}
  \caption{\label{tab:convergence_p1}$L^2$-error of linear ($k=1$), quadratic ($k=2$), and cubic ($k=3$) iso-parametric finite elements for the vector Helmholtz equation with experimental order of convergence (eoc) in the grid width $h$ that tends towards $2$, $3$, and $4$, respectively.}
\end{table}

\begin{figure}[ht]
  \centering
  \begin{subfigure}[b]{0.25\textwidth}
    \centering
    \includegraphics[width=\textwidth]{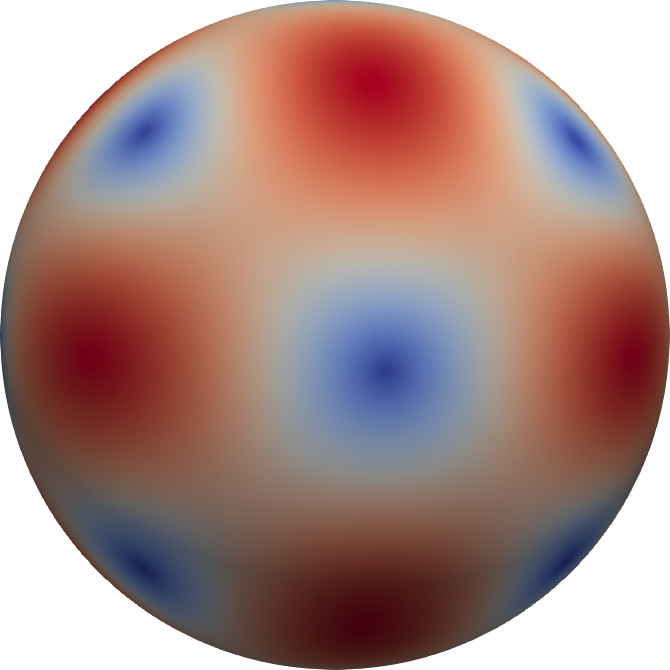}
    \caption{\label{fig:vector_helmholtz:sphere}Sphere}
  \end{subfigure}
  \hspace{2cm}
  \begin{subfigure}[b]{0.25\textwidth}
    \centering
    \includegraphics[width=\textwidth]{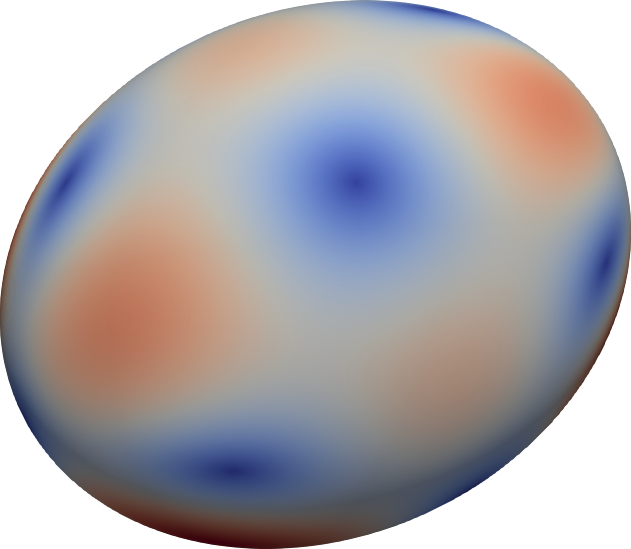}
    \caption{\label{fig:vector_helmholtz:ellipsoid}Ellipsoid}
  \end{subfigure}
  \caption{\label{fig:vector_helmholtz}Solution of the vector Helmholtz equation on the sphere and on the ellipsoid.}
\end{figure}

\subsection{Moving Grids}
Let $\Gamma_h\subset\mathbb{R}^3$ be a smooth closed and stationary reference surface. A map $\X\,:\,\Gamma_h\times[0,T]\to\mathbb{R}^3$  then defines a parametrization of a family of surfaces $\Gamma(t)\subset\mathbb{R}^3$ over this reference manifold:
\begin{equation}
  \Gamma(t) = \left\{\X(x,t)\,:\,x\in\Gamma_h\right\}\,.
\end{equation}
The evolution of this family of surfaces is characterized by its velocity $v(X,t)\in\mathbb{R}^3$ at each point $X=X(x,t)\in\Gamma(t)$,
\begin{equation}
  \partial_t X(x,t) = v(X(x,t),t)\,.
\end{equation}

We consider the surface evolution driven by its mean curvature, the geometric \emph{mean curvature flow} flow, see \cite{DeDzEl2005,KoLiLu2019,Dz1990}. Therefore, we introduce $H \coloneqq \operatorname{tr}(\kappa)$ the mean curvature of the surface with extended Weingarten map $\kappa=-n\otimes \nabla_\Gamma$, and the surface evolution $v = -H n$.

Utilizing the geometric identity $\Delta_\Gamma X = -H n$, see, e.g., \cite{Dz1990,DzEl2013}, a weak formulation of the evolution law can be written:
For all $t\in [0,T]$, find $\X(\cdot,t)\in [H^1(\Gamma_h)]^3$ such that
\begin{equation}
  \int_{\Gamma(t)} \partial_t x(t)\cdot y\,\textrm{d}\Gamma(t) = -\int_{\Gamma(t)}\sum_{i}\triangledown_{\Gamma(t)}x^i(t)\cdot\triangledown_{\Gamma(t)}y^i\,\textrm{d}\Gamma(t)\quad\forall y\in [H^1(\Gamma(t))]^3\,,
\end{equation}
with the surface identity $x(t)=X(\cdot,t)$. Note that on the right-hand side of that equation, we find the componentwise surface gradient of the parametrization.

In order to discretize this equation, we introduce a splitting of the time-interval $[0,T]$ into discrete time steps $0<t_0<t_1<\ldots<t_N=T$ with generic time step width $\tau=t_s - t_{s-1}$ and denote by $\X_s\cong\X(\cdot,t_s)$ the parametrization at time step $t_s$. Correspondingly, we denote by $\Gamma_s=\X_s(\Gamma_h)$ the surface at that time step. Since the grid-function $\X_s$ is parametrized over the reference surface $\Gamma_h$ we replace the integration over $\Gamma_s$ by an integration over the reference surface using a transformation of the surface elements $\text{d}\Gamma_h\to\text{d}\Gamma_s$.

For the discretization in space, we introduce the finite-element space $V_h^r$ of Lagrange finite-elements on $\Gamma_h$,
\[
  V_{h}^r\coloneqq \big\{v\in C^0(\Gamma_h)\,:\,v\circ\X_e\in\mathbb{P}_r\,\forall e\in\mathcal{G}_h\big\}\,.
\]
Then we get the discrete variational formulation by simple Euler discretization in time:
Let $\X_0$ be a given initial parametrization. For all $s=1,\ldots,N$, find $\X_s\in [V_h^r]^3$ such that
\begin{equation}
  \int_{\Gamma_h} (\X_s - \X_{s-1})\cdot \mathbf{Y}\,\textrm{d}\Gamma_{s-1} = -\int_{\Gamma_h}\tau\sum_{i}\triangledown_{\Gamma_{s-1}}X_t^i\cdot\triangledown_{\Gamma_{s-1}}Y^i\,\textrm{d}\Gamma_{s-1}\quad\forall \mathbf{Y}\in [V_h^r]^3\,.
\end{equation}

So, while traversing the reference grid, we need the geometry of the curved grid from the last time step $\Gamma_{s-1}=X_{s-1}(\Gamma_h)$. This is given by the grid-function $\X_{s-1}$:

Initially we construct a \texttt{DiscreteGridViewFunction}:
\begin{c++}
auto X = discreteGridViewFunction(refGrid->leafGridView(), order);
auto X_e = localFunction(X);

// interpolate the initial surface parametrization
auto perturbedSphere = [](auto const& x) { return ...; };
Functions::interpolate(X.basis(), X.coefficients(), perturbedSphere);
\end{c++}

This grid-function additionally provides a global basis that can be localized to an element:
\begin{c++}
auto localView = X.basis().localView();

// traverse the reference grid
for (const auto& e : elements(X.basis().gridView()))
{
  // bind the local function to the element
  X_e.bind(e);

  // bind the localized basis to the element
  localView.bind(e);

  // the localized basis provides a local finite-element
  auto const& localFE = localView.tree().child(0).finiteElement();
  auto const& localBasis = localFE.localBasis();

  // ...
}
\end{c++}

On each element of the reference grid, we can construct a curved geometry. This can be used to obtain the integration element and the transform of the local gradients of the local basis-functions to the actual domain of the curved element.
\begin{c++}
  LocalFunctionGeometry geometry(referenceElement(e), X_e);

  const auto& quadRule = QuadratureRules<double,2>::rule(e.type(), quad_order);
  for (const auto& qp : quadRule) {
    // integration element dG_{s-1}
    double dS = geometry.integrationElement(qp.position()) * qp.weight();

    // the inverse of the transposed geometry Jacobian
    auto Jtinv = geometry.jacobianInverseTransposed(qp.position());

    // evaluate the local basis Jacobians in the quadrature point
    std::vector<FieldMatrix<double,1,2>> shapeGradients;
    localBasis.evaluateJacobian(qp.position(), shapeGradients);

    // transform the local basis Jacobians to the real element
    std::vector<FieldVector<double,3>> gradients(shapeGradients.size());
    for (std::size_t i = 0; i < gradients.size(); ++i)
      Jtinv.mv(shapeGradients[i][0], gradients[i]);
  }
\end{c++}

\begin{figure}[ht]
  \begin{center}
    \includegraphics[width=.25\textwidth]{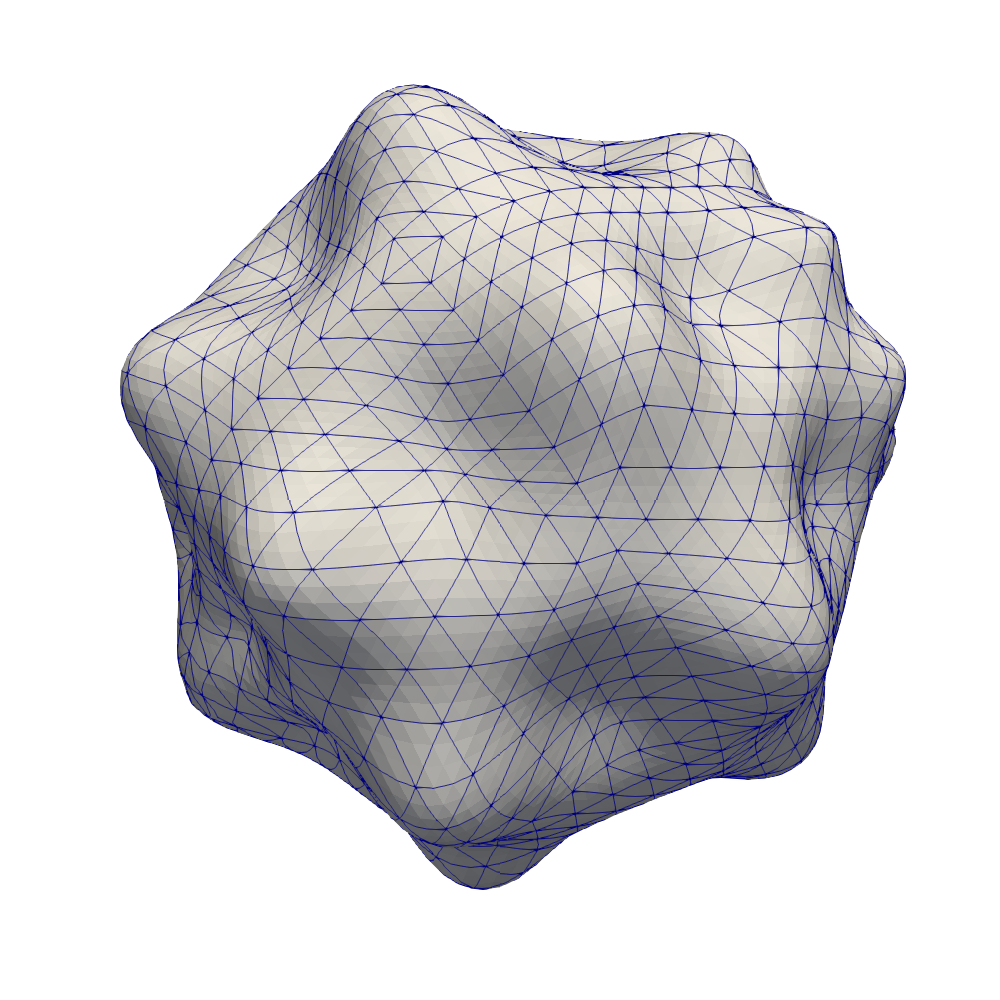}%
    \includegraphics[width=.25\textwidth]{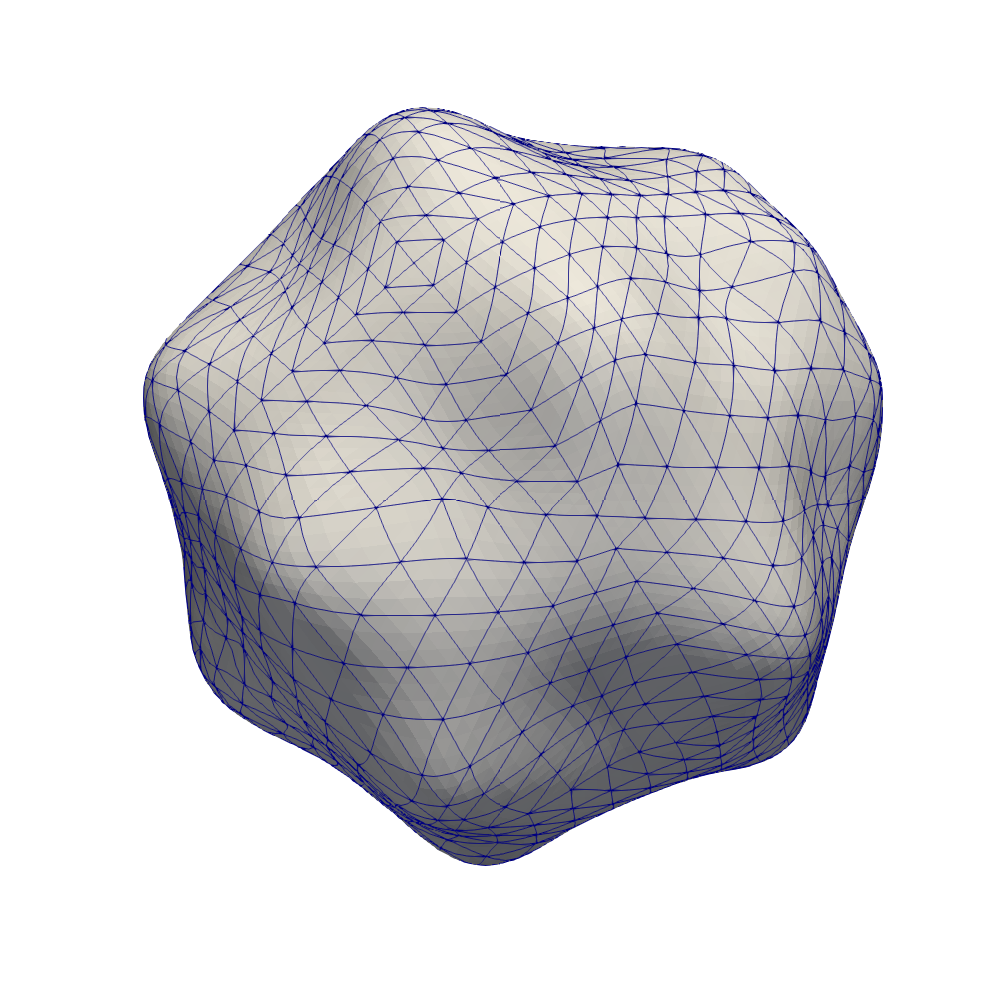}%
    \includegraphics[width=.25\textwidth]{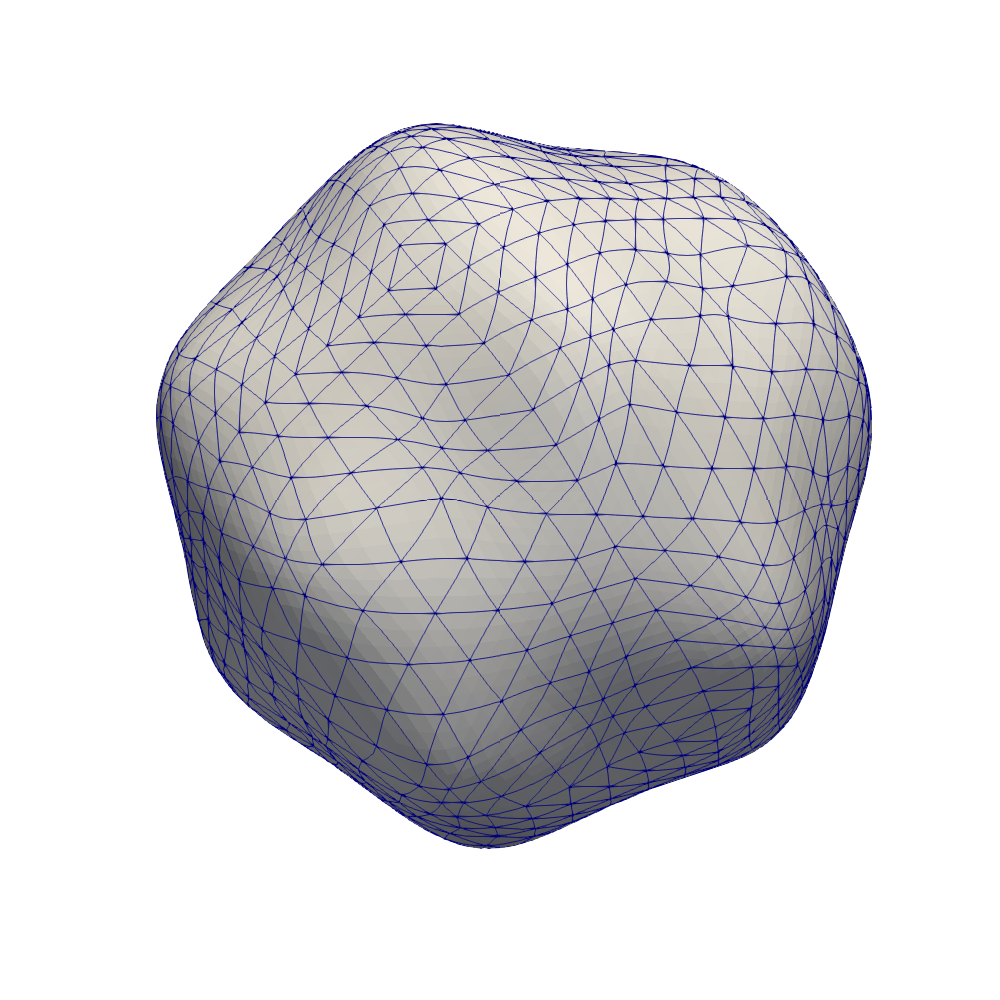}%
    \includegraphics[width=.25\textwidth]{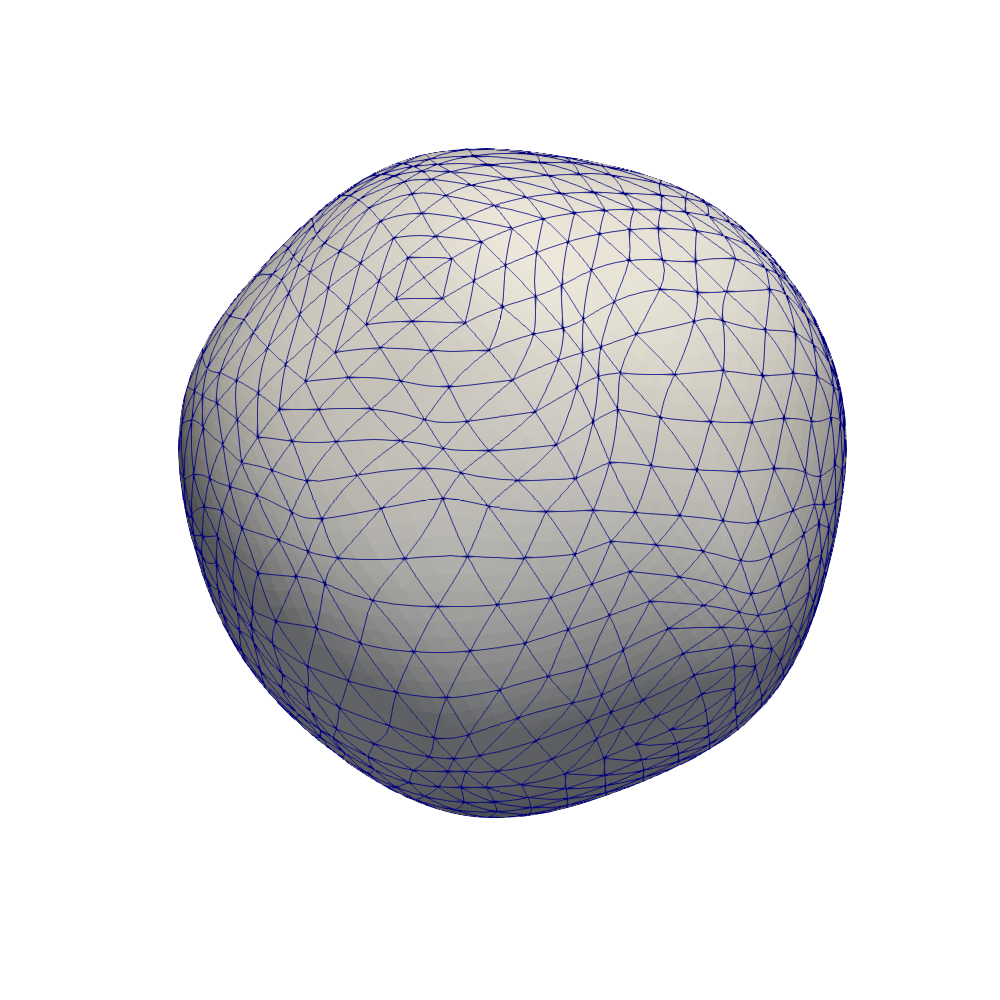}%
  \end{center}
  \caption{\label{fig:mcf}Mean-curvature flow of a perturbed spherical surface with parametrization of polynomial order 2 at four different time-steps in the evolution.}
\end{figure}

The evolution of a perturbed initial sphere can be found in \autoref{fig:mcf}. It starts the evolution by smoothing high curvature regions while continuously shrinking the surface. Eventually the surface gets sphere-like with a radius that tends to zero.

\section{Conclusion and Outlook}\label{sec:conclusion_outlook}
We have implemented parametrized and curved geometries for the \Dune\ framework by wrapping grid-functions or differentiable functions into the \texttt{Geometry} interface defined by the \textsc{Dune-Grid} module. Additionally, we have implemented a wrapper for flat grids providing a curved geometry on traversal, while preserving the grid topology and element connectivity.

It is shown in several examples how these wrappers provide high flexibility while preserving simple usage patterns. In a numerical study we have verified the implementation by showing classical geometry error bounds to be achieved.

The \Dune\ modules \textsc{Dune-CurvedGeometry} and \textsc{Dune-CurvedGrid} provide not only the geometry and grid wrappers but also utility functions to simplify the work with curved geometries. These utilities include some reference geometries and surface projections, as well as grid-functions for various requirements. A long-term goal is to integrate these \Dune\ modules into the core functionality of \Dune.

The implementation of the \Dune\ modules is neither restricted to only surface parametrizations nor to grids without boundaries. The transformation of 1d, 2d, or 3d geometries is implemented and handling of boundary parametrized intersections is included. A future work is the development and application of more grid-functions for the geometry parametrization, e.g., based on a b-spline basis or p-adaptive Lobatto functions. The latter may allow for boundary adapted high-order parametrizations with inner elements described as affine mappings. Another topic for further studies is the differentiability of the element-functions used in the grid parametrization. Automatic or numeric differentiation of the projections $\X$ or $\X_e$, as well as implicit differentiation of the levelset function $\psi$, could be a way to allow for exact geometry parametrizations of more surfaces.

\section*{Acknowledgments}
This work was supported by German Research Foundation (DFG), Research Unit \emph{Vector- and Tensor-Valued Surface PDEs} (FOR 3013)

\appendix
\section{Appendix}

\subsection{Input and output of curved geometries}\label{sec:input_output}
In all the examples above, a reference grid $\mathcal{G}_h$ is provided by reading a surface mesh from file. \Dune\ provides a multitude of grid file readers, but lacks support for a reader that can read curved geometries directly. This would allow to not only start from a reference grid with an analytical projection, but to provide a discrete representation of the curved surface from the beginning. Many meshing tools allow to directly construct such curved meshes and provide a file format that is able to represent the additional nodes for a parametrization. We have implemented two readers, the \texttt{VtkReader} for the VTK file format and a \texttt{Gmsh4Reader} for the \Gmsh\ file format. Both associated meshing and visualization tools, ParaView, \cite{ParaView2005}, and \Gmsh, \cite{GeRe2009}, allow to design curved geometries and to export these in the mentioned file formats.

In addition to file readers, the result of a numerical computation must be exported to allow visualization and postprocessing. Our tool of choice is ParaView, supporting the VTK file format also for curved geometries. We have implemented a grid and data writer for this file format.

The module \textsc{dune-vtk} provides grid readers and writers with flexible input and output policies in the VTK file format, while the module \textsc{dune-gmsh4} provides grid readers for the Gmsh4 file format. In the following code snippets we show the VTK reader and writer, a corresponding Gmsh reader works analogously.

\subsubsection{File Readers for Curved Grids}
When reading a higher-order grid from file, we need to construct both the reference grid and the parametrization. The reference grid could be obtained by evaluating the higher-order grid representation in the element's corner vertices, whereas the parametrization must be extracted from the additional Lagrange nodes stored in the file.

In order to read these nodes and to construct an element connectivity, an input policy called \textit{grid-creator} must be provided. It creates local vertex coordinates and element indices from the fields read from file and passes those to a \texttt{GridFactory} to create the actual grid. The grid-creator for VTK and Gmsh4 files with parametrized grid elements is called \texttt{LagrangeGridCreator}.

The grid itself does not contain the additional Lagrange nodes, but a parametrization or coordinate mapping describing the higher-order geometries. Thus, one needs to associate these nodes to a local Lagrange basis. We provide a grid-function representation of the higher-order geometries parametrized over the extracted reference grid. This grid-function is represented by the \textit{grid-creator} itself.

\begin{c++}
// <dune/vtk/vtkreader.hh>
// <dune/vtk/gridcreators/lagrangegridcreator.hh>

using Grid = FoamGrid<2,3>;
using Creator = Vtk::LagrangeGridCreator<Grid>;
auto grid = VtkReader<Grid, Creator>::createGridFromFile("filename.vtu");
\end{c++}

To extract the parametrization in addition to the reference grid, we need to obtain the grid-creator directly, that acts as a grid-function after reading from the file:

\begin{c++}
using Grid = FoamGrid<2,3>;
GridFactory<Grid> factory;
Vtk::LagrangeGridCreator creator{factory};
VtkReader reader{creator};
reader.read("filename.vtu");

// construct the reference grid
auto grid = factory.createGrid();
\end{c++}

Thus, the grid-function can be used to fill any other storage to represent the geometry, e.g., by interpolating into a \texttt{DiscreteGridViewFunction}, or can be used directly for the parametrization of the \texttt{CurvedGrid}:

\begin{c++}
CurvedGrid curvedGrid{*grid, creator, creator.order()};
\end{c++}

\subsubsection{VTK Writer}
VTK supports higher order cell types including Lagrange parametrizations of cells since version 9. This allows to directly write the curved geometries to files. A corresponding output policy, called \emph{data-collector}, is added to support these cell types. This data-collector is responsible for transforming a grid-view into a list of point coordinates and a connectivity table. Additionally, it collects values associated to the point coordinates, if data should be written to the VTK file.

For writing higher-order Lagrange parametrized grids, in addition to the corner vertices of grid elements internal Lagrange nodes are written to the file. The connectivity table collects these nodes in a specific order so that they can be associated to Lagrange basis functions. The corresponding data-collector is called \texttt{LagrangeDataCollector}:

\begin{c++}
// <dune/vtk/datacollectors/lagrangedatacollector.hh>

namespace Vtk {
  template <class GridView, int ORDER = -1>
  class LagrangeDataCollector;
}
\end{c++}

The template parameter \texttt{GridView} represents the type of the grid-view that shall be written and the second (optional) parameter \texttt{ORDER} represents the Lagrange polynomial order of the cell parametrization. This second parameter is optional since also runtime polynomial order is supported, by passing the polynomial order parameter to the constructor instead.

\begin{c++}
LagrangeDataCollector (const GridView& gridView, int order = ORDER)
\end{c++}

In case no constructor parameter for the order is given, the template \texttt{ORDER} parameter is used as default value. Note that either in the template parameter or in the constructor argument a positive value for order must be given.

The corresponding writer object can be instantiated by either passing a data-collector object or by letting the writer construct it with the given grid-view object.

\begin{c++}
using DataCollector = Vtk::LagrangeDataCollector<GridView,4>;
using Writer = VtkUnstructuredGridWriter<GridView, DataCollector>;

// a) default construct the data-collector with the passed gridView
Writer vtkWriter1{gridView};

// b) construct the data-collector before and pass it to the writer
DataCollector dataCollector{gridView};
Writer vtkWriter2(dataCollector);
\end{c++}

Note, we are using an unstructured-grid writer to generate a \texttt{.vtu} file that represents the grid.

\begin{figure}[ht]
  \begin{center}
  \includegraphics[width=.3\textwidth]{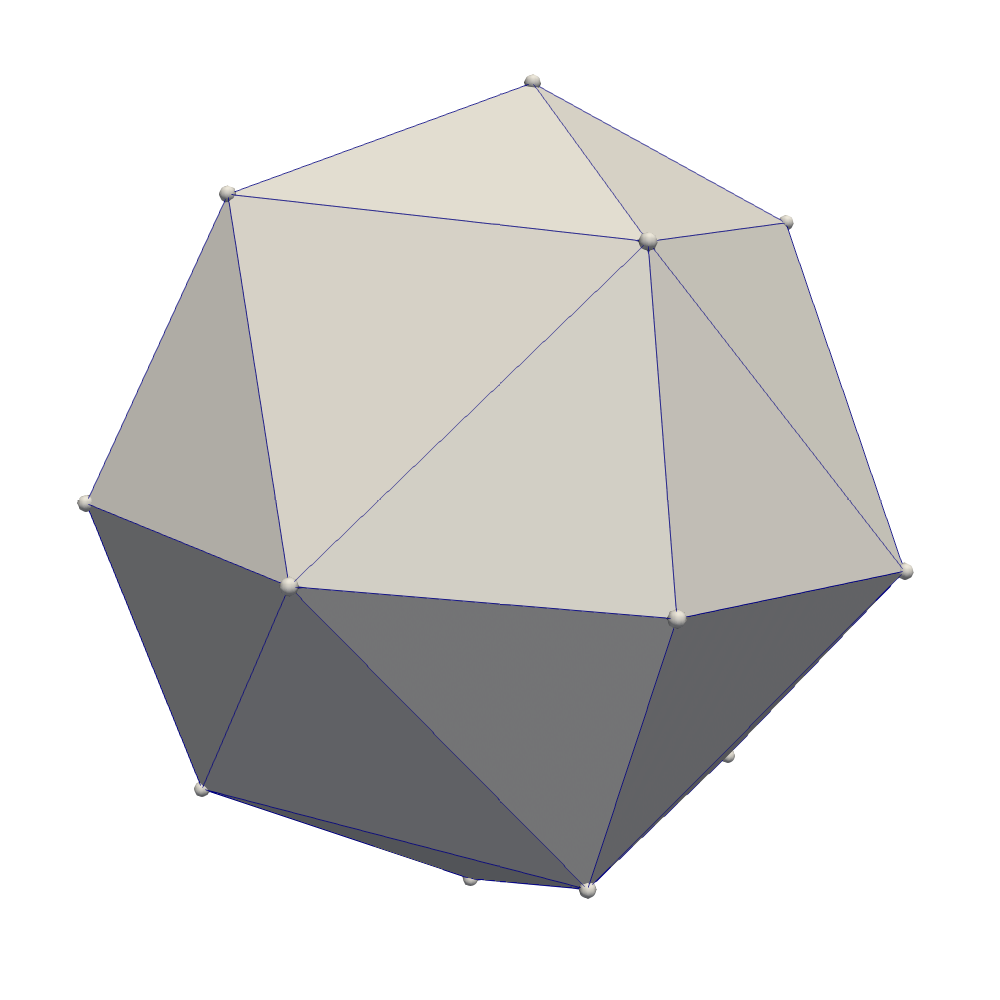}
  \includegraphics[width=.3\textwidth]{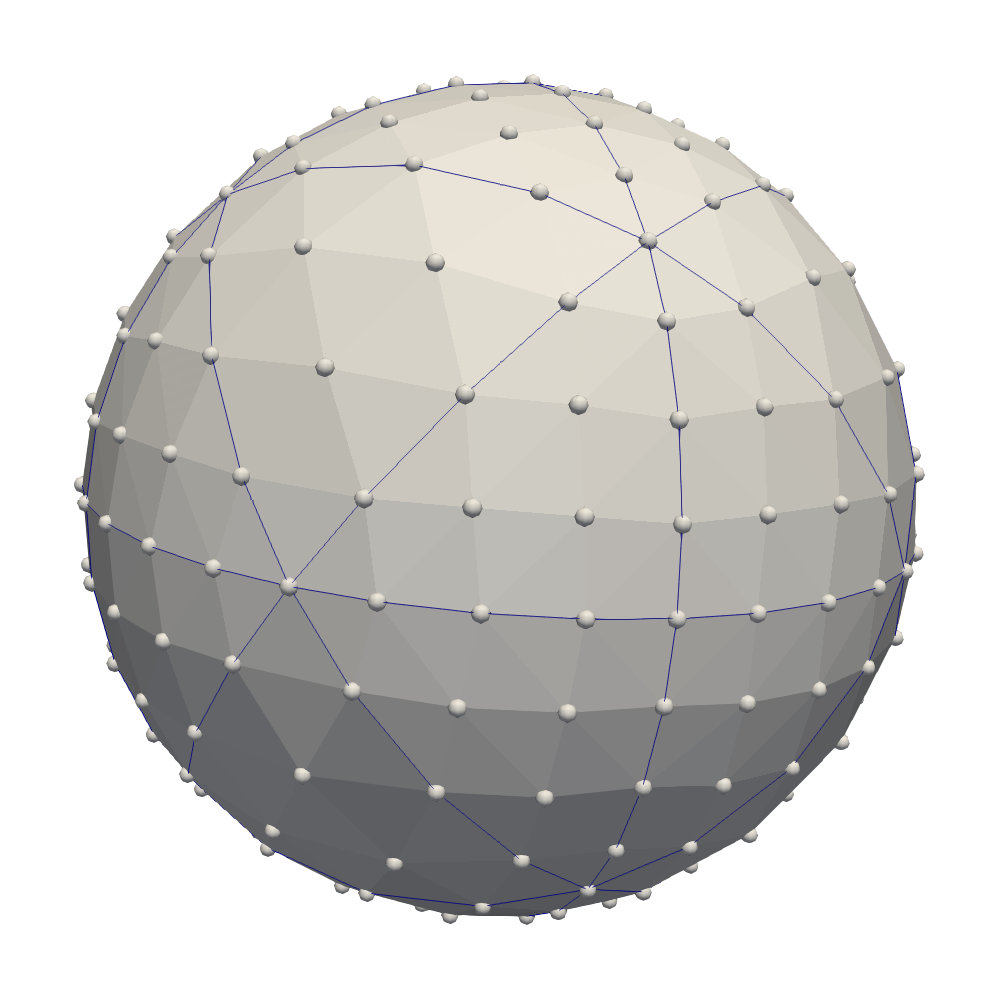}
  \includegraphics[width=.3\textwidth]{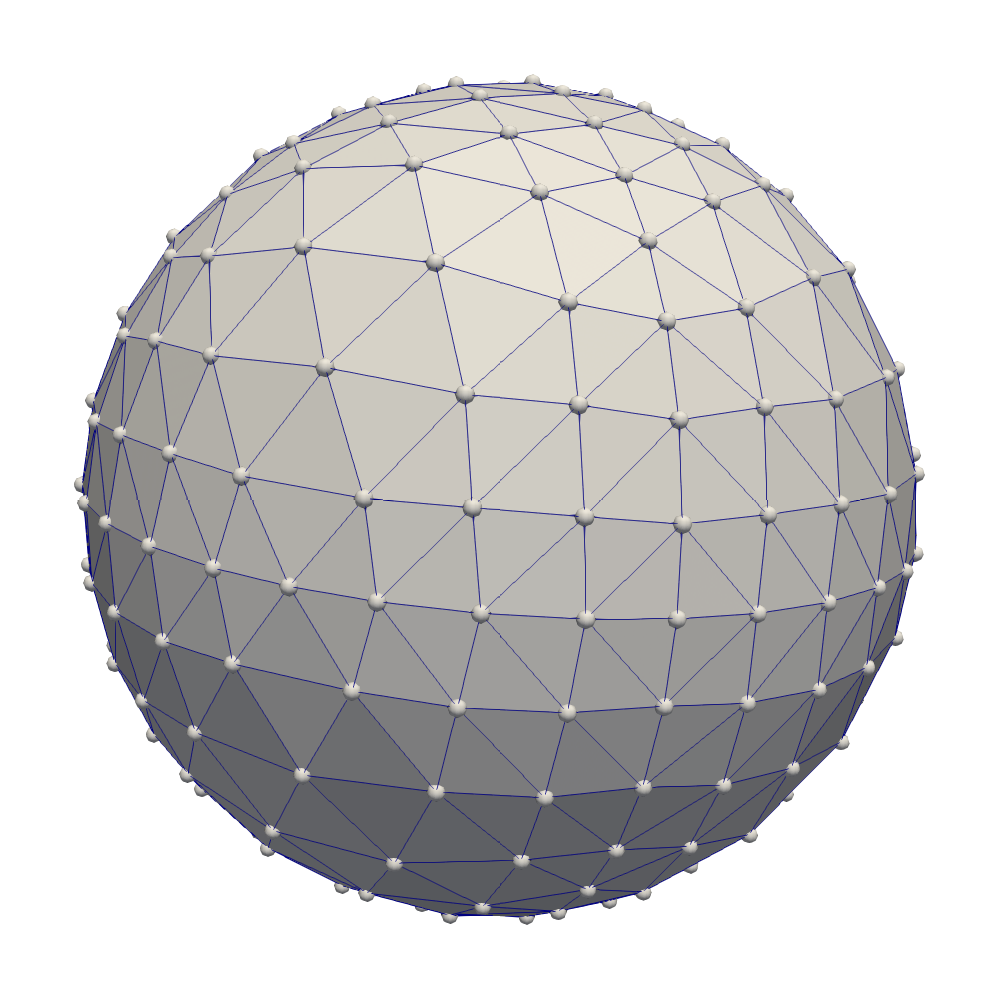}
  \end{center}
  \caption{\label{fig:vtk_visualization_sphere}Three different approximations of the sphere visualized using the VTK writer with ParaView. Shown are the element edges and the Lagrange nodes. Left: reference grid, Center: Lagrange parametrization with polynomial order $k=4$, Right: Lagrange parametrization with polynomial order $k=1$ and two grid refinements.}
\end{figure}

\bibliographystyle{habbrvnat}
\bibliography{bibliography}
\end{document}